\documentclass[useAMS,usenatbib]{mn2e}

\def\var{\hbox{OGLE05155332--6925581}}
\usepackage{graphicx}
 \usepackage{times}

\title[A clue for  Double Periodic Variables]
  {The eclipsing LMC star OGLE05155332--6925581: a clue for Double Periodic Variables}

\author[Mennickent et al.]
  {R.E. Mennickent,$^1$\thanks{E--mail: rmennick@astro--udec.cl.
  Based on observations carried out at ESO and CTIO telescopes: ESO proposal 076.D--0126(A) 
  and NOAO proposal CHILE03B--01}
  Z. Ko{\l}aczkowski$^{1,2}$, G.~Michalska$^2$, G.~Pietrzy\'nski$^{1,3}$, R.~Gallardo$^1$,
  \newauthor
L. Cidale$^{4}$, A. Granada$^{4}$, W.~Gieren$^1$  \\
  $^1$Universidad de Concepci\'on, Departamento de F\'{\i}sica,
      Casilla 160--C, Concepci\'on, Chile\\
  $^2$Instytut Astronomiczny Uniwersytetu Wroclawskiego, Kopernika 11, 51--622 Wroclaw, Poland \\    
  $^3$Warsaw University Observatory, Al. Ujazdowskie 4,00--478, Warsaw, Poland \\
  $^4$ Fac. de Cs. Astron\'omicas y Geof\'isicas, U. Nac. de La Plata and Instituto de Astrof\'isica La Plata (IALP), CONICET, Paseo del Bosque S/N, 1900, La Plata, Argentina.
  }
\date{}

\pagerange{\pageref{firstpage}--\pageref{lastpage}} \pubyear{2007}

\def\LaTeX{L\kern-.36em\raise.3ex\hbox{a}\kern-.15em
    T\kern-.1667em\lower.7ex\hbox{E}\kern-.125emX}

\begin{document}

\label{firstpage}

\maketitle

\begin{abstract}

 We investigate the nature of  \var, one of the brightest members of the enigmatic
group of Double Periodic Variables (DPVs) recently found in the Magellanic Clouds.
The modeling of archival orbital light curves (LCs),  along with the analysis of the radial velocities suggest  that this object is a semi--detached binary with the less massive star transferring matter to the more massive and less evolved star, in an Algol--like configuration. 
We find evidence for additional orbital variability and H$\alpha$ emission, likely caused by an accretion disc around the primary star.
As in the case of  $\beta\,Lyr$  the circumprimary disc seems to be more luminous than the primary, but we do not detect orbital period changes. 
We find that the LC follows a loop in the color--magnitude  diagram during the long cycle; the system is redder when brighter and the rising phase is bluer than during decline. Infrared excess is also present. The source of the long--term periodicity is not eclipsed, indicating its circumbinary origin.  Strong asymmetries, discrete absorption components (DACs)  and a $\gamma$ shift are new and essential observational properties in the infrared H\,I lines. The DACs strength and RV follow a saw--teeth pattern during the orbital cycle.  
We suggest that the system experiences supercycles
of mass outflow feeding a circumbinary disc. Mass exchange and mass loss could produce comparable but opposite effects in the orbital period on a long time scale, 
resulting in a quasi--constancy of this parameter.

\end{abstract}


\section{Introduction: about DPVs and \var}

The star \var~ (MACHO IDs 79.5739.5807 and 78.5739.78; OGLE LMC--SC8--125836) is a member of the group of close binaries named Double Periodic Variables (Mennickent et al. 2003, Mennickent et al. 2005a). These stars show two closely related photometric periodicities; the shorter one reflects the binary period but the cause for the longer periodicity (or quasi--periodicity since it is not exactly cyclic, Mennickent et al. 2005b) still is unknown. We have selected \var~ for a detailed monitoring and study since it is  relatively bright, has a large amplitude of the long term changes and its eclipsing nature could help to shed light on the nature of the phenomenon of Double Periodic Variables, especially their evolutionary stage and long--term variability.

~\var~ shows a long term periodicity of 188 day and an eclipsing variability with orbital period 7.2843 days (Mennickent et al. 2003).  The standard multi--color photometry of this system is available in several catalogues published recently.
Zaritsky et al. (2004) gives simultaneous $U$ = 15.040 $\pm$ 0.032, $B$ =  15.575  $\pm$ 0.031 and $V$ = 15.587 $\pm$ 0.057 mag. Wyrzykowski et al. (2003) gives $V$ = 15.38, $B$ = 15.32 and $I$ = 15.27 mag at maximum. 
Ita et al (2004) provides single--epoch simultaneous three color infrared imaging with $J$ = 15.586, $H$ = 15.566 and $K_{s}$ = 15.490 mag. No epoch is available for these measurements. \var~ is one of the brightest DPVs in the $LMC$, in consequence the quality of the available photometry  is good enough for concluding about physical properties of this object. On the other hand, their photometric credentials are typical for DPVs so it can be considered as a representative case.

~\var ~ appears in the field as a visual double, with a nearby companion 3.8\arcsec to the East.  
 This neighboring star was also observed photometrically and spectroscopically. The OGLE position for 
the companion is RA = 05:15:53.88 and DEC = $-$69:25:58.1 and the photometry gives $V$ = 16.947, ($B-V$) = 1.812 and
($V-I$) = 1.821 mag. The $I$--band light curve (LC) consisting of 395 datapoints gives a mean of $I$ = 15.072 mag
with $RMS$ 0.006 mag. 2MASS data for this star give $J$ = 13.928
$\pm$ 0.048, $H$ = 13.046 $\pm$ 0.061 and $K$ = 12.817 $\pm$ 0.055 mag. A finding chart based on an OGLE image
taken at the $I$--band is given in Fig.\,1.

We provide a description of our observations and archive photometric data in Section 2 along with our methods
of data reduction and analysis. The results of our LC modeling and spectroscopic study 
are given in Section 3 and a discussion is presented in Section 4. Our conclusions are given in Section 5.

 In this paper we will label with subindex 1 those parameters of  the more massive star (the primary) and with 2 those of less massive star of the binary pair.

\begin{figure}
\scalebox{1}[1]{\includegraphics[angle=0,width=8.5cm]{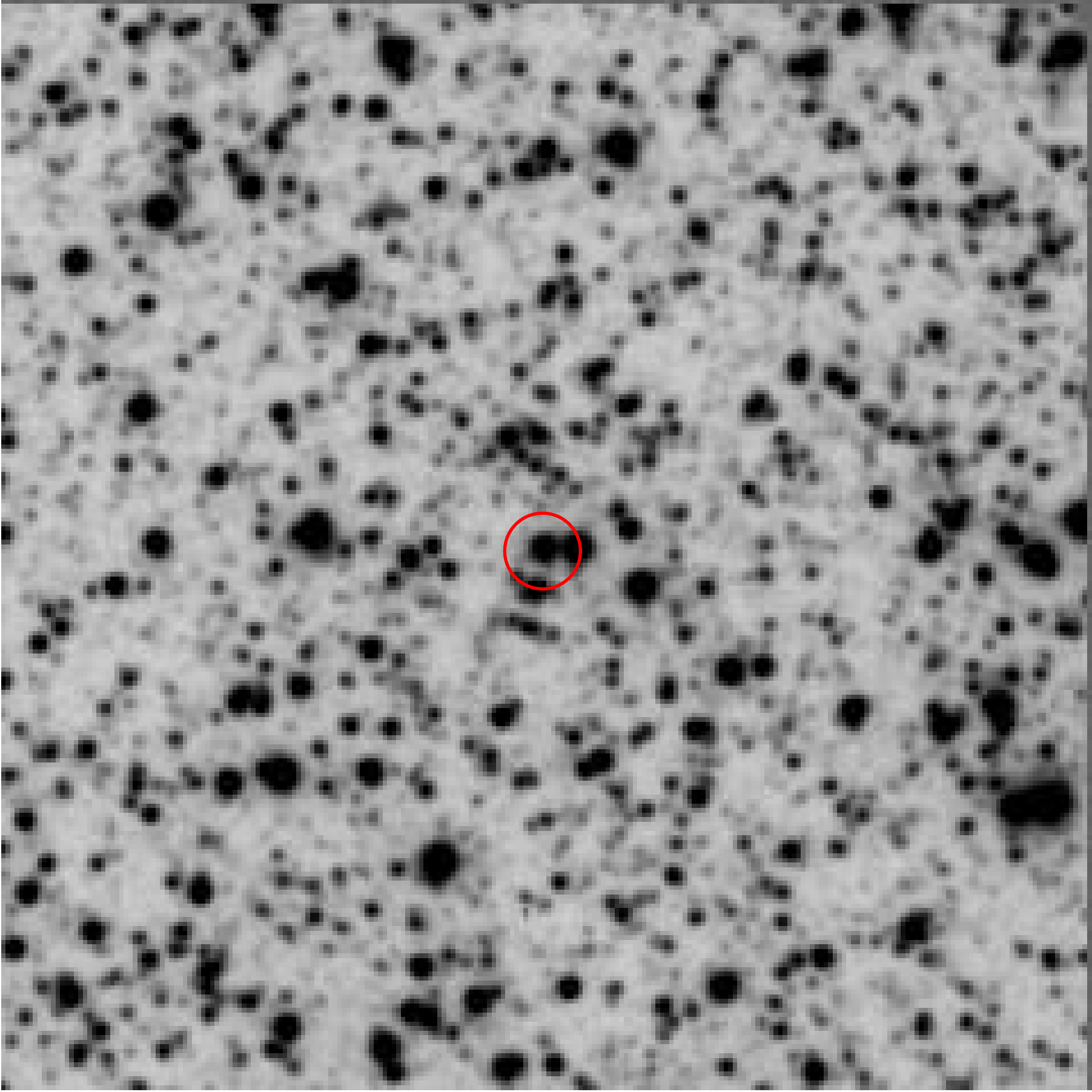}}
 \caption{The finding chart of \var~  based on an I--band OGLE image. North is up and east right. The field of view  is 1\farcm7 $\times$ 1\farcm7 
 and centered at RA(2000) = 05:15:53.32 and DEC(2000) = $-$69:25:58.1. The circle indicates the position of \var.}
  \label{1}
\end{figure}

\begin{figure}
\scalebox{1}[1]{\includegraphics[angle=0,width=9.3cm]{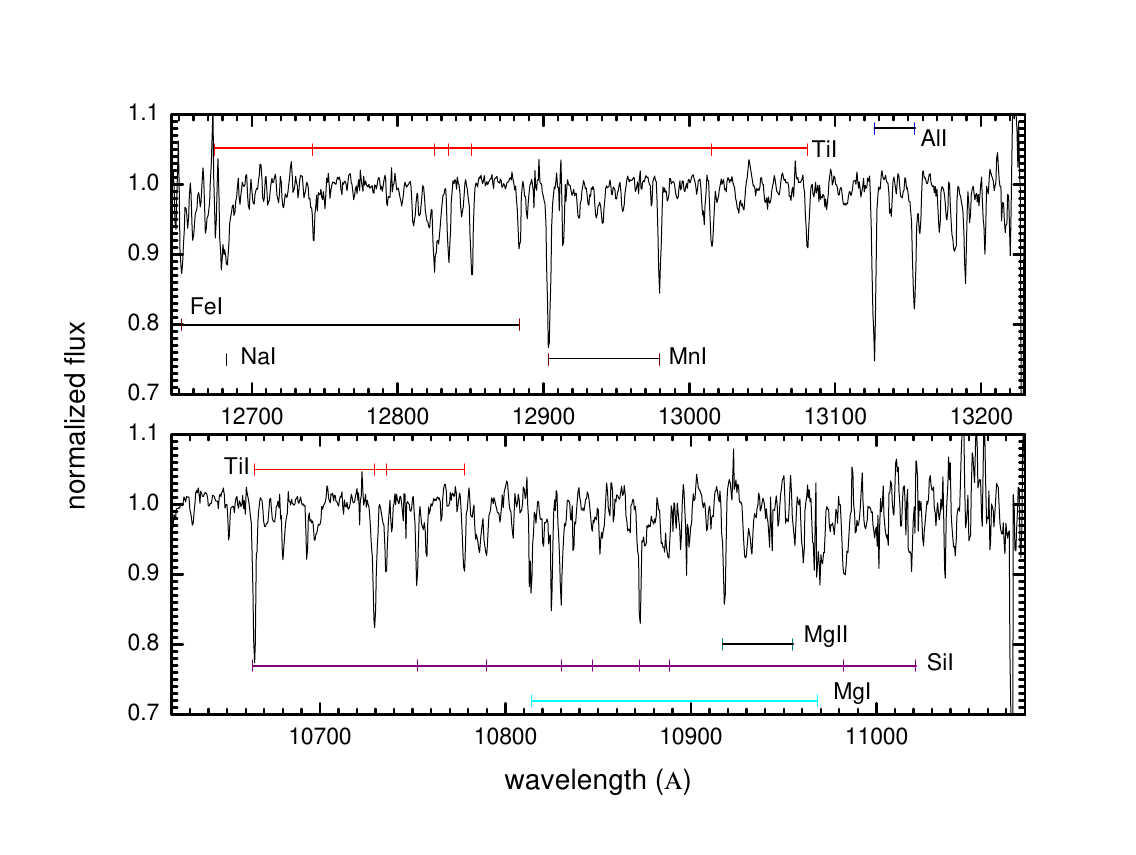}}
 \caption{The average infrared spectra for the nearby companion observed in the same slit. Equivalent total exposure time is 3.3 hours.
 The spectra have been corrected by the stellar radial velocity. Line identification is based on the work by Wallace et al. (2000).}
  \label{2}
\end{figure}

\begin{figure*}
\scalebox{.9}[.9]{\includegraphics[angle=0,width=18cm]{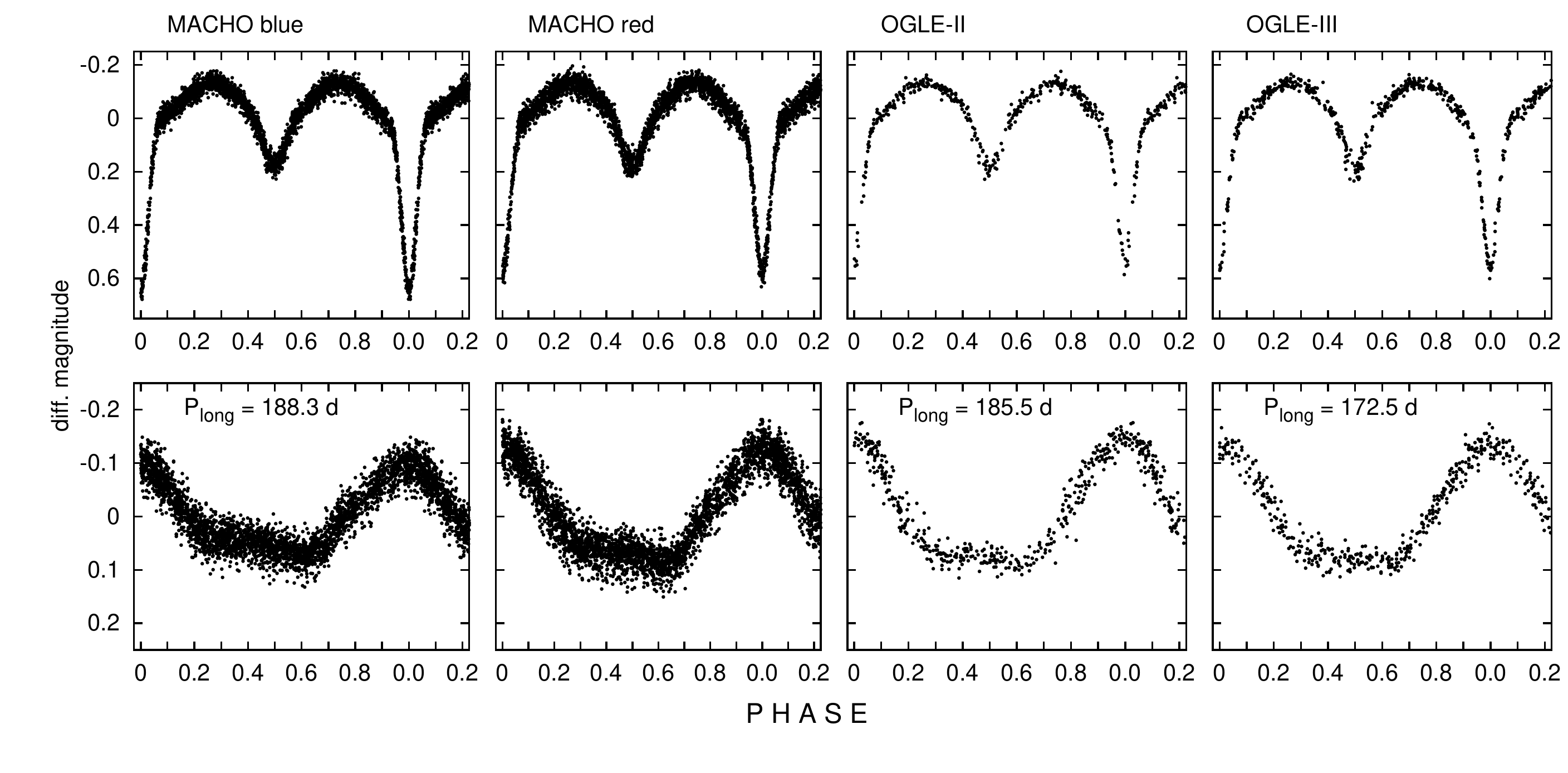}}
 \caption{Results of the Fourier light curve decomposition showing the orbital (upper graphs) and long--term (lower graphs)  variability separately.}
  \label{3}
\end{figure*}

\section{Spectroscopic observations and sources of photometry}

We have monitored \var~ spectroscopically in the optical wavelength range
during August 2003 and February 2004, and in the infrared region during December 2005 and December 2006 and also obtained an optical high
resolution spectrum at a single epoch in 2007.
In this section we give details of these observations. In addition, we have used public domain photometric data
available from the MACHO (Allsman \& Axelrod 2001) and OGLE\,II databases (Udalski et al. 1997, Szyma\'nski 2005) along with OGLE\,III data. The MACHO light curves analyzed in this paper consist of 1469 photometric measurements obtained at the "red" filter (central wavelength 7000 \AA) and 1555 measurements obtained at the "blue" filter (central wavelength 5200 \AA) during HJD\,2\,448\,826.136 and 2\,451\,536.193 (red) and 2\,451\,541.999 (blue). We also analyzed 896 OGLE\,II and III magnitudes obtained at the I--band (central wavelength 8000 \AA) spanning a time interval between HJD 2\,450\,457.649 and 2\,453\,823.498. The total time covered by the photometric data analyzed in this paper is 13.7 years.

\begin{table}
\centering
 \caption{ The radial velocities for \var~ in km s$^{-1}$ referred to the Local Standard of Rest. 
 HJD' means HJD$-$2\,450\,000.  The orbital and superphases are given along with detection $d$ of
 He\,I\,4471 (p: positive, w: weak, w: weak doubtful and n: negative).}
 \begin{tabular}{@{}cccccccc@{}}
  \hline
UT--date & HJD' &$\Phi_{o}$ &$\Phi_{s}$ & H$\delta$ &H$\gamma$ &H$\beta$ &$d$\\
\hline
Aug08  &   2859.8912 &0.582 &0.074&     226 &   154 &  208 &n  \\
Aug20  &   2871.8437 &0.222 &0.144&      $-$ &     $-$ &   $ -$ &n  \\
Sep03  &   2885.7998 &0.138 &0.225&     293 &   308 &  328 &n  \\
Sep11  &   2893.7845 &0.235 &0.271&     366 &   410 &  357 &n  \\
Sep24  &   2906.8188 &0.024 &0.346&    359 &   366 &  312 &w  \\
Sep26  &   2908.8315 &0.300 &0.358&    508 &   295 &  459 &p  \\
Oct10  &   2922.8309 &0.222 &0.439&    408 &   430 &  316 &w  \\
Oct22  &   2934.7777 &0.862 &0.508&    207 &   212 &  185 &p  \\
Oct24  &   2936.7097 &0.127 &0.520&     142:&   320 &  240 &w  \\
Nov27  &   2970.8199 &0.810 &0.717&    132 &   229 &  215 &w: \\
Nov28  &   2971.7277 &0.935 &0.723&    225 &   191 &  231 &p  \\
Dec10  &   2983.7592 &0.586 &0.792&     101 &   136 &  244 &p  \\
Dec11  &   2984.7069 &0.716 &0.798&   205 &   103 &  241 &n  \\
Dec12  &   2985.6706 &0.849 &0.803&     219 &   164 &  165 &n  \\
Dec17  &   2990.7294 &0.543 &0.833&     217 &   306 &  274 &n  \\
Dec18  &   2991.7520 &0.684 &0.839&    235 &   269 &  148 &w: \\
Dec19  &   2992.7313 &0.818 &0.844&  109 &   166 &  155 &w: \\
Dec30  &   3003.6639 &0.319 &0.908&    346 &   378 &  397 &n  \\
Dec31  &   3004.6364 &0.452 &0.913&    290 &   274 &  248 &w: \\
Jan11  &   3015.6778 &0.968 &0.977&     267 &   268 &  241 &n  \\
Jan12  &   3016.6215 &0.098 &0.983&       $-$ &   467 &  448 &n  \\
Jan20  &   3024.7157 &0.209 &0.030&  391 &   451 &  471 &n  \\
Jan21  &   3025.6757 &0.341 &0.035&    391 &   267 &  405 &p  \\
Jan22  &   3026.6209 &0.471 &0.041&    256 &   253 &  347 &p  \\
Jan31  &   3035.6290 &0.707 &0.093&    116 &   182 &   95 &p  \\
Feb01  &   3036.6371 &0.846 &0.099&    164 &   67  &  131 &p  \\          
\hline
\end{tabular}
\end{table}

\subsection{Time resolved optical spectroscopy}

26 spectra of \var~ in the wavelength region of 3900$-$5300 \AA~were obtained with the
1.5m  telescope of the Cerro Tololo Interamerican Observatory (CTIO), Chile,  
in service mode using the Cassegrain Spectrograph with grating 26 tilted 15\fdg94. 
The slit width of 1\arcsec yielded a nominal resolution of 2 \AA. 
He--Ar comparison spectra obtained along every science exposure 
provided wavelength calibration functions with typical $RMS$ of 0.1 \AA~ (6 km s$^{-1}$ at  H$\beta$).
 Standard techniques of spectrum extraction, calibration and analysis were applied using 
IRAF\footnote{IRAF is distributed by the National Optical Astronomy Observatories,
    which are operated by the Association of Universities for Research
    in Astronomy, Inc., under cooperative agreement with the National
    Science Foundation.}  tasks. 
We measured the radial velocity (RV) of the Balmer lines with simple gaussian fits. 
The average RV of the standard star LTT\,2415, measured every observing night, 
resulted to be 255 $\pm$ 36 km s$^{-1}$, which compares well with the value of 253.4 km s$^{-1}$
reported in the SIMBAD database. Our measured velocities for \var~ are given in Table 1. 

\subsection{The MIKE high resolution spectrum}

We obtained one spectrum of \var~ with  the Magellan Inamori Kyocera Echelle (MIKE)  in the Baade Magellan telescope of
Las Campanas Observatory, Chile, the night of February 23 (UT), 2003 (UT--start 00:58:39, HJD at mid--exposure 
2\,452\,693.55208). 
We used a 1800 sec exposure and a slit width of 0.35\arcsec, 
resulting in a resolving power $\approx$ 70\,000.  Our spectrum was taken at 
orbital phase 0.756 and superphase 0.110 (see below).
 The calibration and extraction of the spectrum were performed with  standard IRAF tasks.
The spectrum was wavelength calibrated with several hundreds comparison lines of ThAr spectra taken before 
and after the exposure. The $RMS$ of the calibration function was about $10^{-3}$ \AA.

\subsection{J--band infrared spectroscopy with the ESO ISAAC}


Our observations were conducted at the  European Southern Observatory (ESO) Paranal observatory,  Chile, with  the
Infrared Spectrometer and Array Camera (ISAAC) in the SW--MR mode in service mode (http://www.eso.org/instruments/isaac/). The detector was the Hawaii Rockwell array with
nominal gain 4.5 e/ADU and readout noise 11 electrons. This spectroscopic mode provided
a plate scale of 0.147\arcsec per pixel. We used a slit width of 1\arcsec yielding a spectral resolution of 
2.4 \AA~ and resolving power $R \approx$ 4550. 
We observed \var~ at three different spectral regions: the j region
between 1.061$-$1.108 $\mu$m, the sz region between 1.154$-$1.215 $\mu$m and the jplus region between 1.265$-$1.324 $\mu$m. The observations were done as usual in the infrared, with the technique of dithering.
This method consists in observing the object at two or more positions along the
slit. The sky is effectively removed by subtracting one frame from the other, registering the two
beams and then subtracting again. 
 The first steps of image reductions like geometrical distortion correction, 
flat fielding and bias removal were done by the ISAAC pipeline software at the observatory. 
The wavelength calibration was done identifying
night sky lines in the image and building a wavelength calibration function for every spectrum.
Extraction was done with the usual IRAF tasks. 

We used a generic template of infrared atmospheric lines
to remove telluric lines. This template was rebinned and resolution degraded to match the
characteristics of our spectra. 
Then we divided the science spectra by the telluric spectrum after finding the best 
wavelength shift and vertical scaling providing the best telluric feature removal. 
Equivalent widths  of the absorption lines were measured by line fitting with Lorentz functions. For those lines showing discrete absorptions in their wings, we deblended the profile with two Lorentz functions of variable depth and width to take into account the blending. For comparison, we also measured the equivalent width with the standard formula integrating the line flux under the interpolated continuum, yielding results about 15\% lower than those obtained with the first method. For the
OB205093 spectrum, only this method was possible due to the large line asymmetry. We measured radial velocities in the infrared spectra determining the wavelength at minimum  of the flux. 
We determined an internal error of 10 km s$^{-1}$ from radial velocities of the comparison star, red companion described in Section 1, placed in the same slit. Details of the observations are given in Table 2. We found that the comparison star is a probable member of the LMC, its RV referred to the LSR is +246 $\pm$ 10 km s$^{-1}$. The spectra for this star are plotted in Fig.\,2, along with line identifications; the star could be considered as a flux and radial velocity standard in the LMC. Using the
ratio between the Mn\,I feature strength to that of Mg\,I in the formula provided by Wallace (2000) for Galactic giant stars 
we obtain a temperature of  3953 $K$. 
We also note that the Hipparcos ISAAC telluric standards Hip\,026368 and Hip\,029635 are Be stars; they show double peak Pa$\beta$ emission with peak separation of 205 and 235 km s$^{-1}$, respectively. The first star is the well  known Be star AZ\,Dor (B9.5\,Ve) and the second one (HD 44533, B8\,V) is a new Be star.

 \begin{table}
 \centering
\begin{minipage}{90mm}
 \caption{Observing log for the infrared spectroscopy. We give the exposure time in seconds  $\Delta t$,
 the observing block name, the seeing value as measured in the acquisition image (s1) and read it from the image
 header (s2), in arcseconds, and the spectral band. NA means not available.}
 \begin{tabular}{@{}cccccc@{}}
 \hline
date(UT)    &   $\Delta t$/4 & MJD &OB&s1$-$s2 &band  \\  
\hline   
21--22/12/05 &   600  &   3726.2744-3726.2958 &051221&  0.64$-$0.93& j       \\
10--11/01/06 &   600  &   3746.1699-3746.1913 &205091&  0.54$-$0.68& j       \\
10--11/01/06 &   600  &   3746.1992-3746.2205 &205094&  NA$-$0.69& sz        \\
11--12/01/06 &   500  &   3747.1895-3747.2074 &205088&  0.58$-$0.79& jplus       \\
11--12/01/06 &   600  &   3747.2141-3747.2355 &205092&  NA$-$0.56& j      \\
15--16/02/06 &   600  &   3782.1062-3782.1276 &205095&  0.60$-$0.69& sz       \\
16--17/02/06 &   600  &   3783.0996-3783.1209 &205097&  0.57$-$0.92& jplus       \\
14--15/04/06 &   600  &   3839.9825-3840.0038 &205093&  0.69$-$0.78& j           \\
17--18/03/06 &   500  &   3811.9970-3812.0150 &205090&  0.60$-$0.83& j       \\
12--13/10/06 &   600  &   4021.2740-4021.2954 &205098&  0.75$-$1.32& jplus       \\
09--10/11/06 &   600  &   4049.3222-4049.3435 &205099&  0.56$-$0.45& jplus       \\
06--07/12/06 &   600  &   4076.2686-4076.2899 &255548&  0.66$-$1.36& jplus       \\
\hline
\end{tabular}
\end{minipage}
\end{table}

\begin{table}
 \caption{Periods and amplitudes found in the light curves.}
 \begin{tabular}{@{}rccrcc@{}}
 \hline
 period (d) & range (mag) &MACHO &period (d) &range (mag) &OGLE  \\  
\hline   
7.28428&    0.773 & blue &7.28425&    0.660 &  II\\
  188.4&    0.166 &  blue&  185.6&    0.240 &  II\\
7.28431&    0.713 & red &7.28430&    0.651 & III\\
  188.3&    0.206 &  red& 172.5&    0.220 &  III\\
\hline
\end{tabular}
\end{table}

\begin{table*}
\centering
 \caption{ The best system parameters ({\it p}),  derived assuming a stellar origin for the deeper eclipse. At every band we give the flux fraction $l_{1}/(l_{1}+1_{2})$ at maximum.  $g$ is the surface gravity in cm s$^{-2}$. We also give the parameters for the best fit RV = $\gamma + K_{2} sin(2\pi\Phi_{o}$). $f$ is the mass function. For our calculations we have used published values of $\mu$ = 18.39 and $E(B-V)$ = 0.17.}
 \begin{tabular}{@{\, \vline \,}ccc@{\, \vline \,}cc@{\, \vline \,}cccc@{\, \vline \,}}
 \hline                             
\multicolumn{1}{c}{$p$} &
\multicolumn{1}{c}{MACHO blue} &
\multicolumn{1}{c}{OGLE\,II} &
\multicolumn{1}{c}{$p$} &
\multicolumn{1}{c}{value} &
\multicolumn{1}{c}{Line} &
\multicolumn{1}{c}{$\gamma$ (km s$^{-1}$)} &
\multicolumn{1}{c}{$K_{2}$ (km s$^{-1}$)} &
\multicolumn{1}{c}{$f (M_{\sun})$} \\
  \hline     
 $q$  &   0.287&0.393 &     $Sp1$   &?        &Pa$\beta$ &347(5) &167(10) &3.52(63)              \\ 
   $\log g_{1}$ &3.96&3.79&    $Sp2$ &B6\,III           &Pa$\gamma$ &380(18) &152(25) &2.65(1.30)    \\
 $\log g_{2}$ & 2.86&2.88&   $M_{V,2}$ &$-$2.83         &H$\beta$ &280(12) &122(16) &1.37(54)  \\
 $T_{1} (K)$ & 25\,000 & 25\,000 &  $M_{bol,2}$     &$-$4.03  & H$\gamma$&271(13) &123(18) & 1.40(61)  \\
 $T_{2} (K)$  &   14\,090&14\,330&  $R_{2}/a$& 0.29  &H$\delta$ &277(11) &134(15) & 1.82(61)  \\
    $i (\deg)$ &    80.4&80.4&$R_{1}/a$&0.16  & & & &   \\
$f_{B}$ &   0.419&0.419 &$P_{o} (d)$&7.284297(10)   & & & &      \\
$f_{V}$  &0.392&0.418 & $P_{s} (d)$&188.2$-$172.5 & & & &  \\
 $f_{I}$ &0.360 &0.387& && & & &  \\
  $f_{J}$ &0.324&0.351 & && & & &   \\
     $f_{H}$ & 0.315&0.342&  & & & & &  \\
  $f_{K}$&0.308&0.335& &     & & & &  \\
 \hline
\end{tabular}
\end{table*}

\begin{table}
\centering
 \caption{Equivalent widths and radial velocities in the infrared spectra. $\Phi_{o}$ is the orbital
 phase.}
 \begin{tabular}{@{}ccccc@{}}
 \hline
OB    &$\Phi_{o}$&EW(\AA) & RV(km s$^{-1}$) &line  \\  
\hline   
051221&0.599  &3.5 &330 &Pa$\gamma$\\
205091&0.331  &5.1 &535 &Pa$\gamma$\\
205094&0.336  &$-$   &$-$   &$-$\\
205088&  $-$     &5.5 &$-$   &$-$\\
205092&0.475  &5.7 &334 &Pa$\gamma$\\
205095&0.264  &  $-$  &$-$   &$-$\\
205097&0.401  &7.1 &447 &Pa$\beta$\\
205093&0.210  &4.2 &559 &Pa$\gamma$\\
205090&0.368  &4.9 &467 &Pa$\gamma$\\
205098&0.098  &5.8 &461 &Pa$\beta$\\
205099&0.948  &5.9 &279 &Pa$\beta$\\
255548&0.648  &4.6 &229 &Pa$\beta$ \\
\hline
\end{tabular}
\end{table}

\section{Results}

\subsection{Photometry from OGLE and MACHO}

 The light curves were analyzed with a Fourier disentangling method. 
This allowed us the separate the Fourier amplitudes and frequencies of the two different periodic signals detected 
in the  time series photometry. 
A summary of these parameters for \var~ is given in Table 3.
The results of our analysis of the OGLE and MACHO photometry are displayed in Fig.\,3. The light curves are folded
with the orbital and the long--term period. The double eclipsing nature of the variable is evident as well as the
hump--like character of the long--term light curve, whose variability is limited to half a cycle only, and shows larger amplitude at the red band. We observe the eclipses slightly deeper in the blue band. The improved ephemeris for the main minimum is  (the numbers in parentheses are the errors with the zeros omitted):\\

$HJD_{0} = 2\,450\,000.1392(21) + 7.284297(10) \times E$. \hfill(1)\\

\noindent

We note that the long--term  variability source is not eclipsed. This result contradicts the finding by Mennickent et al. (2005a, hereafter M05) based on the visual inspection of the light curves phased with the orbital period for two DPVs (their Fig.\,12). The apparent absence of long--term variability during the main eclipse turned to be only a visual effect produced by the large
magnitude gradients found around minimum. The better treatment of the data by the Fourier  decomposition technique completely separates the short and long--term variability and reveals that the visual impression is not real. Our finding suggests that the source of the long--term variability is placed around the binary system, and this is the rule for all eclipsing DPVs (Ko{\l}aczkowski \& Mennickent in preparation).

In  Fig.\,4 we show the O--C diagram for times of maximum of the long--term variability. 
The maximum moments and their errors were measured at every cycle by fitting the LC with the
average light--curve defined for a given band as a spline function.
There is a clear change in behavior after cycle number six; the long variability becomes faster, and the O--C diagram is consistent with a period change from 188.2 d $\rightarrow$ 172.5 d. Our data does not reveal if this transition is gradual or consist of discrete period jumps.  
The shorter period is observed during the OGLE\,III epoch, and we note in Fig.\,3 that the shape and amplitude of the long--term light curve remains unchanged, besides the notable period change. The ephemeris for  the maximum of the long--term variability during OGLE\,III epochs is $HJD_{max} = 2\,452\,674.6 + 172.5 \times E$. \\

During minimum and maximum of the  long cycle light the system has the most blue and red color, respectively.
Fig.\,5 shows the color--magnitude diagram for the long--term variability of the MACHO red light curve, indicating magnitudes taken during rising or decline from maximum. We observe a clear pattern in the diagram: the loop
characterized by bluer rising than decline phase. 
 This loop is reminiscent of that observed by de Wit et al. (2006) in light curves of outbursting Be stars in the Magellanic Clouds,  and is also a strong evidence for a circumbinary source for the long period variability.

\begin{figure}
\scalebox{1}[1]{\includegraphics[angle=0,width=9cm]{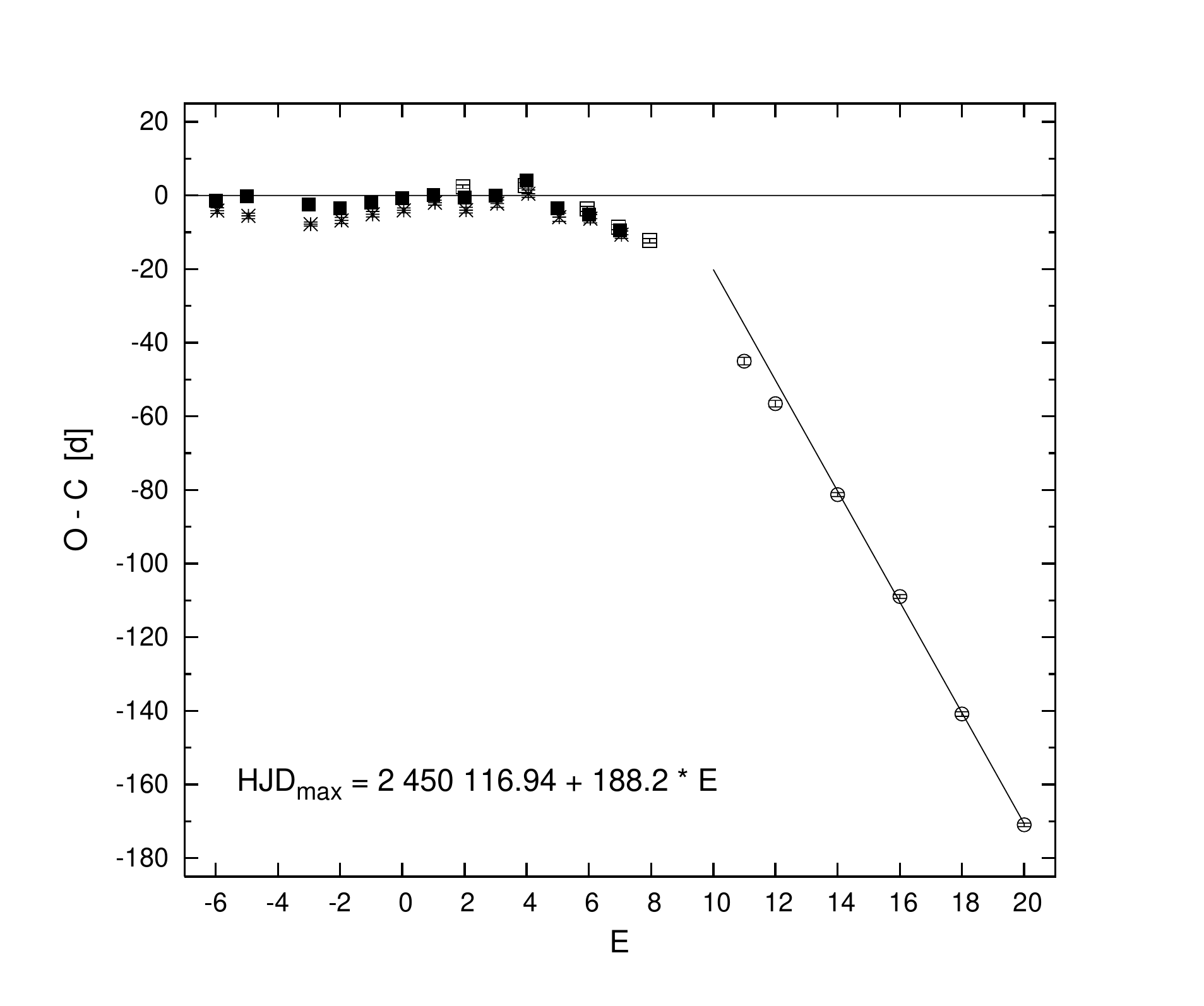}}
 \caption{The O--C diagram for the maxima of the long--term variability. Filled squares correspond to MACHO--red data,
 asterisks to MACHO--blue data, open squares to OGLE\,II data and open circles to OGLE\,III data.}
  \label{4}
\end{figure}

\begin{figure}
\scalebox{1}[1]{\includegraphics[angle=0,width=9cm]{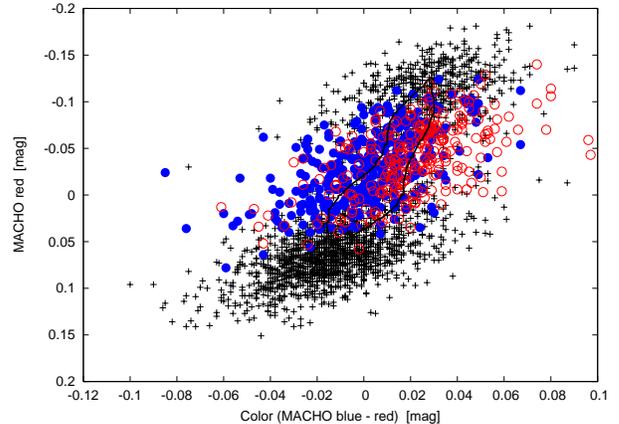}}
 \caption{The color--magnitude diagram for the long--term variability in the MACHO red LC. 
 Solid circles are magnitudes taken during the
 rising and open circles are those obtained during decline from maximum. 
Magnitudes taken at the maximum and broad minimum  are shown by pluses. 
The loop--like solid line follows the average temporal trend.}
  \label{5}
\end{figure}

\begin{figure*}
\scalebox{1}[1]{\includegraphics[angle=0,width=12cm]{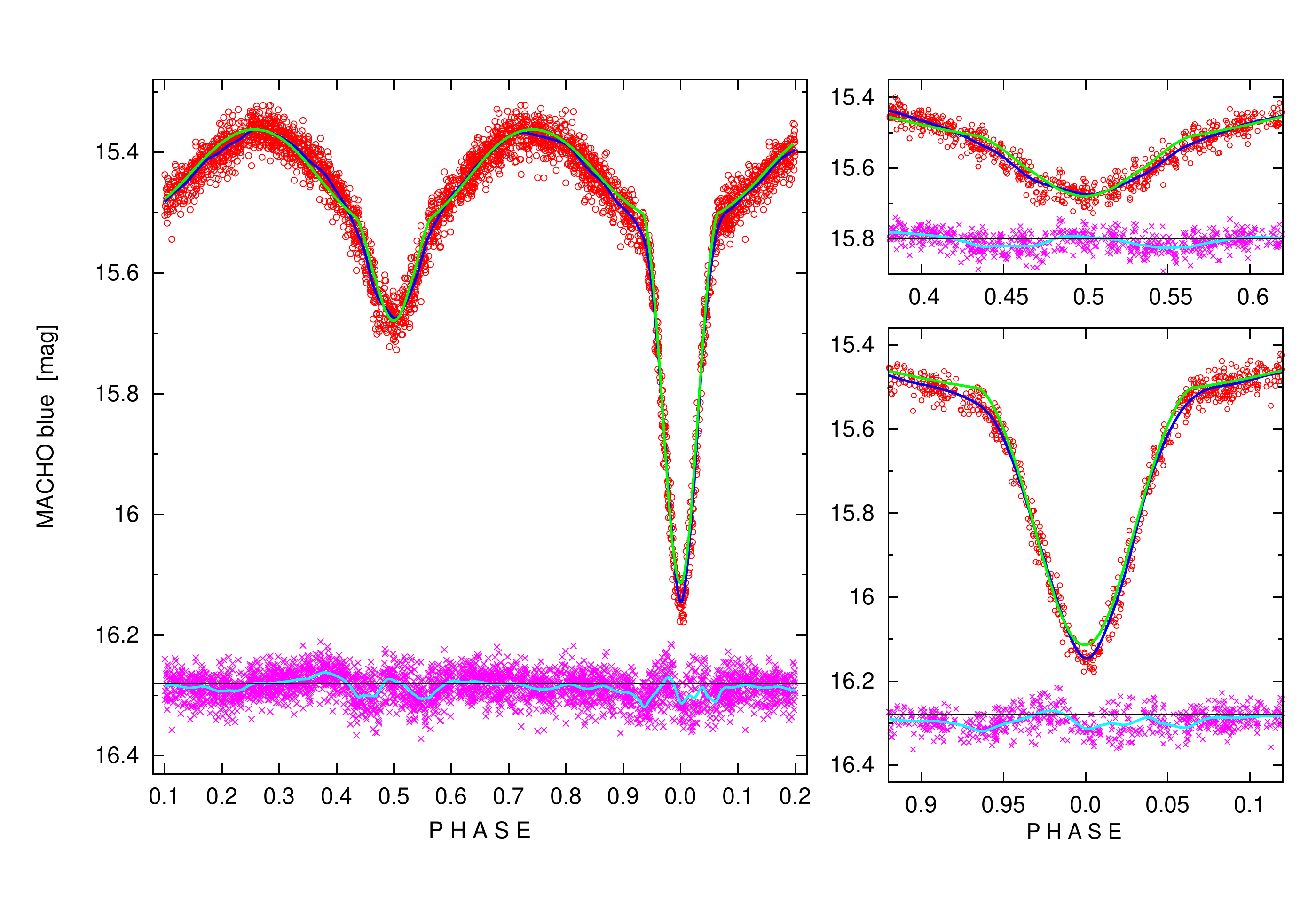}}
 \caption{A detailed comparison of the model and the observations.
Red open circles are observations (MACHO blue; long term variability is removed),
the dark blue line is the fit of spline functions to the observations, the
green line is  the model, based on the parameters of the best solution for the
MACHO blue data. Magenta crosses are the observations minus model, the light
blue line is the spline fit minus model. Zoomed views of eclipses are given 
in the right side panels.}
  \label{6}
\end{figure*}

\begin{figure*}
\scalebox{1}[1]{\includegraphics[angle=0,width=12cm]{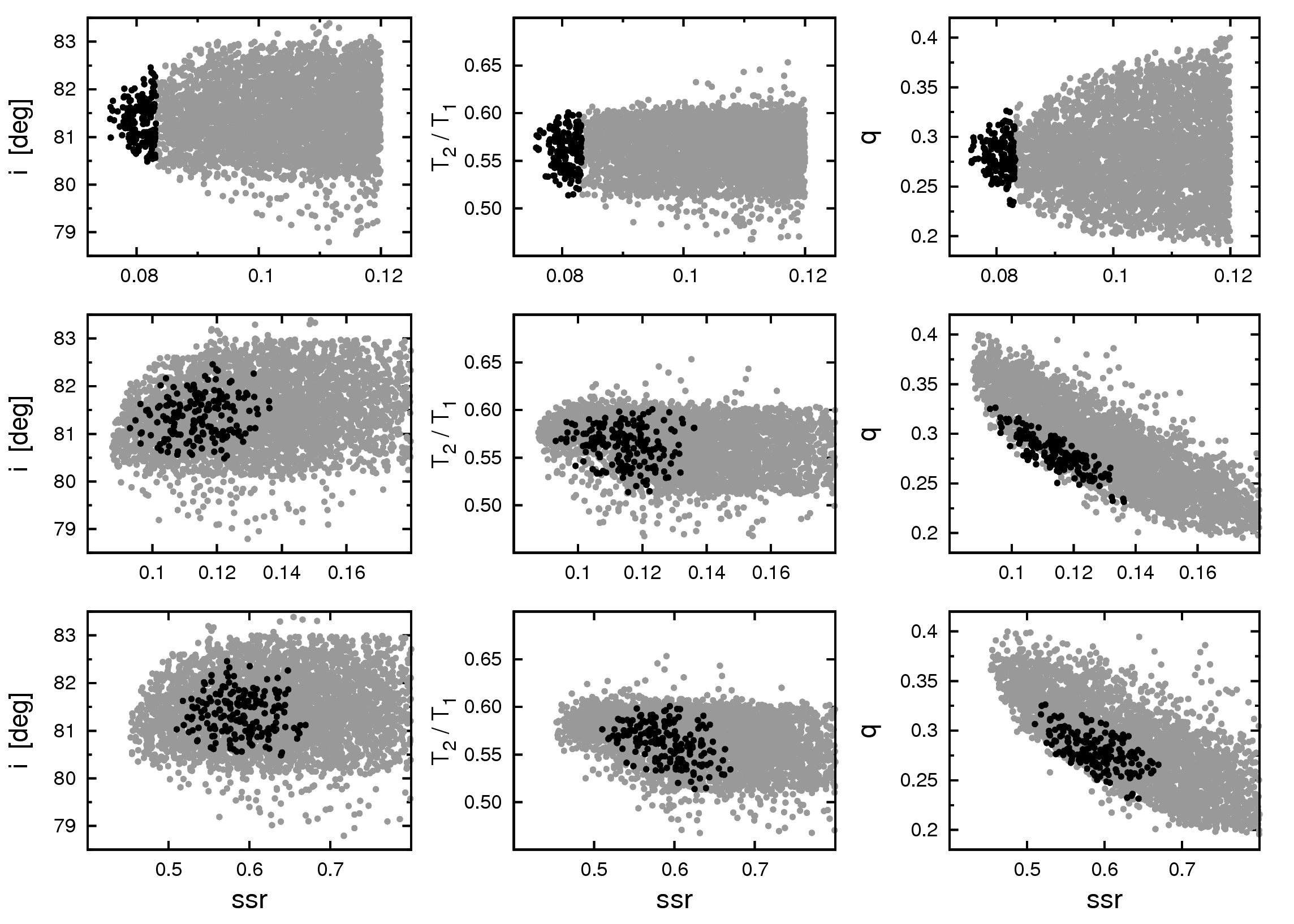}}
 \caption{Results of our Monte Carlo simulations. Every point represents a model, characterized by a set of input parameters for the WD code and its sum of squares of residuals (ssr). The  models in the figures are those minimizing ssr at the MACHO blue LC (upper panels). These models generate LCs at bandpasses corresponding to MACHO red and OGLE LCs. Comparison of these synthetic LCs with the observational data yield to the diagrams of the middle and bottom panels, respectively. Black dots are a 10\% selection of the best models for the MACHO blue LC.}
  \label{1}
\end{figure*}

\begin{figure*}
\scalebox{1}[1]{\includegraphics[angle=0,width=12cm]{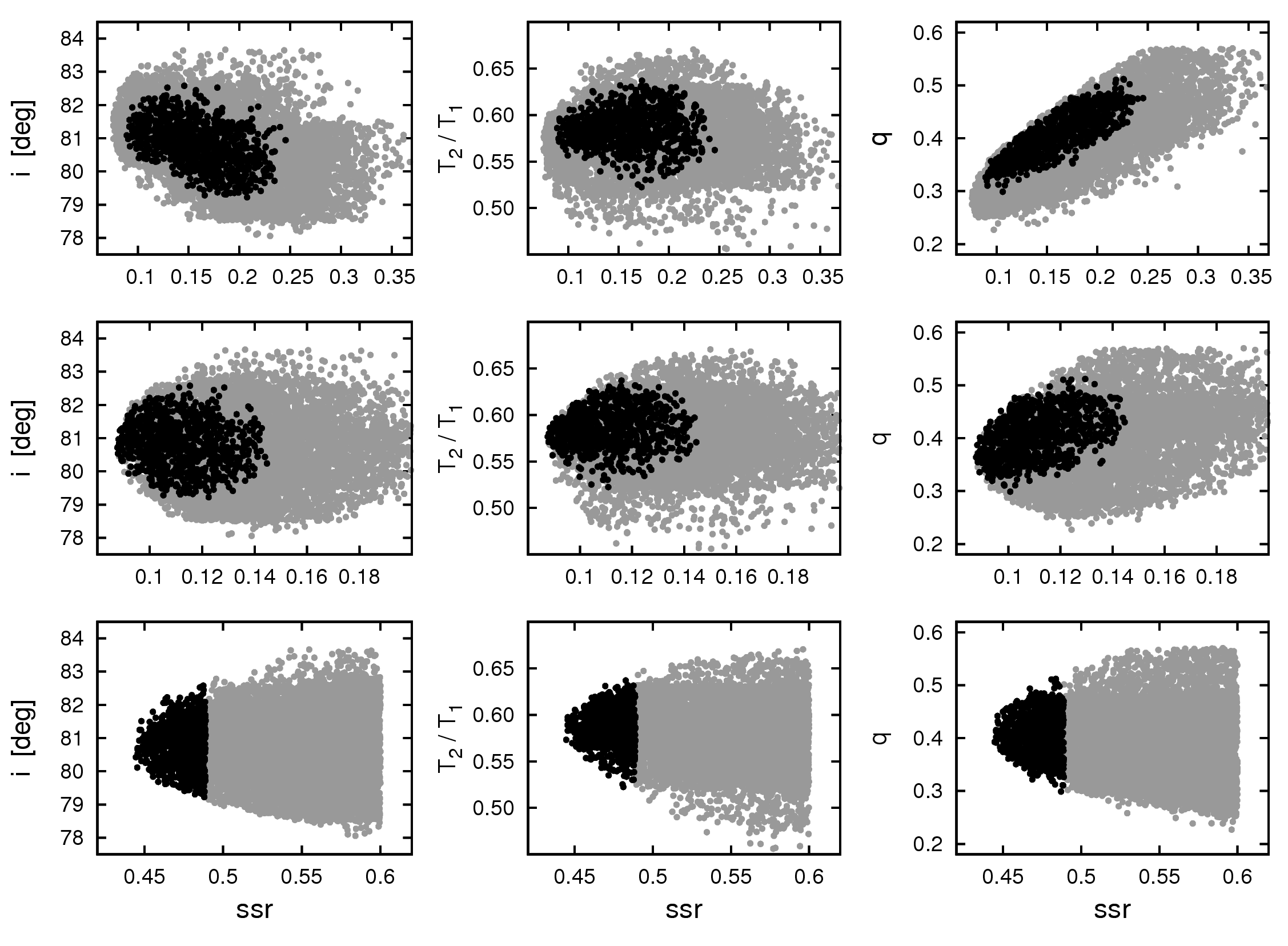}}
 \caption{Same as Fig.\,7 but considering models optimized for the OGLE\,II light curve.}
  \label{8}
\end{figure*}

\subsection{Orbital light curve modeling}

We used the Wilson--Devinney  (Wilson \& Devinney 1971) code for modeling the eclipsing
variability of ~\var~  but we encountered problems with getting convergence of the solution for all available LCs. 
We found also non--physical solutions revealed as   a negative basal constant light.
Therefore we decided to use a light curve modeling program based on WD code that find the best solution mapping a global minimum of summed residuals in a multi parametric space using Monte--Carlo techniques (Michalska \& Pigulski 2005).
We conducted the search for an optimal solution for every band separately
applying also reliable simplifications.
 Circular orbits, synchronous rotation and absence of third light are
assumed by this method. A fixed temperature for the primary component is 
roughly estimated from the brightness and the distance of the system  using the method described
in Michalska \& Pigulski (2005). This iterative method picks up the visual magnitude at maximum and starts assuming equal luminosities for both components and the $T_{1}$ versus $M_{bol}$ relation for Galactic detached main--sequence eclipsing
binaries published by Harmanec (1988). The difference between the Galaxy and
LMC metallicities for the parameters $BC$ and $M_{V}$ was calculated from the models of Bertelli et al (1994). The $T_{1}$ value hence estimated converges to 25\,000 $K$.
The best fit parameters are given in Table 4 and the model light curves 
are shown in Fig.\,6.
Estimates of the fit quality  are given in Figs.\,7 and 8. 
We observe systematic differences between the best solutions at different wavelength bands (Figs.\, 7 and 8). We argue that these differences could be due to the absence of including   a constant basal light (third light, probably residual light from the source of the long--term variability) in our models. In consequence 
we cannot adopt these models as the conclusive physical solution, but
only for a discussion concerning the general properties of the system. 

The potentials of the secondary star in the whole possible range of the mass ratios clearly indicate a semi--detached configuration (Fig.\,9).
 We find $T_{1} = 25\,000 K$, $T_{2} = 14\,090 K, \log g_{1} = 3.96$ ($g$ is the surface gravity in cm s$^{-2}$), $\log g_{2} = 2.86, i = 80\fdg4$ and $q$ = 0.29 from the best
MACHO blue light curve fitting model. These parameters are well confirmed by the analysis of the OGLE\,II light curve except the mass ratio, for which we find $q$ = 0.39. The ratio of the stellar radii $R_{2}/R_{1}$ is in the range 1.79 (for OGLE II--band data) to 1.91 (for MACHO blue).

The temperature and surface gravity of the secondary star agree with the improved values obtained from the spectra reported by M05, viz.\, $T_{2}$ = 13\,500 $\pm$ 1000 $K$, $\log$ g$_{2}$ = 3.2 $\pm$ 0.5 and $M_{v}$  = $-$2.0 $\pm$ 0.3, $M_{bol}$ = $-$2.9 $\pm$ 0.3 and spectral type B5\,IV.

Further detailed comparison between synthetic and observed LCs reveals
additional components (fourth light) modulated with the
orbital cycle of the system.
As we can see in Fig.\,6 the overall model LC is quite good, but some significant deviations are present at phases of both eclipses. The system is fainter than the model at the main minimum and also during ingress and egress of main eclipse. 
In addition, the observed secondary eclipse is wider than the model. These features on the observed LC could be explained by the presence of an optically thick disc of gas around the primary star. This kind of accretion structure could be present in the system due to mass transfer from the secondary to the primary star.
The deeper minimum is the opposite effect that expected in the presence of the third light.
In the observed eclipses both additional components, third and fourth light, should have a contribution
in opposite directions, but in our simple modeling is impossible to estimate their total influence.  We will return to the LC interpretation in the discussion section.

\subsection{The infrared spectra: DACs and $\gamma$ shifts}

We show in Fig.\,10 and 11 the smoothed (7--point average) Pa$\beta$ and Pa$\gamma$ hydrogen lines for \var. Equivalent widths and
radial velocities are given in Table 5.
The H\,I lines appear in absorption with variable asymmetries and depths; the depth is larger at 
phases 0.37$-$0.40 and smaller at phases 0.60$-$0.65. The H\,I profiles looks quite complex in structure.  We identify the main absorption with the secondary star (see Section 3.4) but there is a 
weaker and wider absorption component in the H\,I lines appearing in some phases which we attribute to the primary star.
In Fig.\,12 we show the non smoothed Pa$\beta$ spectrum at phase 0.47, along with a synthetic binary absorption profile built considering the stellar parameters found in Section 3.2. 
In order to reproduce the observed Pa$\beta$ profile, we consider the 
contribution of a primary star with $T_{eff}$ = 25\,000 $K$ and $\log\,g$ = 4 and a secondary 
star with $T_{eff}$ = 14\,000 $K$ and $\log\,g$ = 3. For each object we assumed that the 
photosphere is in radiative and hydrostatic equilibrium and is represented 
by the corresponding density and temperature distributions given by Kurucz's 
(1979) model atmospheres. 
The line profiles are obtained by solving 
the transfer equation and the equations of statistical equilibrium for 
multi--level atoms in the commoving frame with the ETLA code (as described by 
Catala \& Kunasz, 1987). We assumed a spherically symmetric slowly expanding 
atmosphere with a mass loss rate ~ $10^{-9}$ $M_{\sun}$/yr. 
Once the line source function was known, the line profiles for the primary and 
secondary stars were computed in the observer's reference frame and 
 convolved with a grid of rotational profiles. 
 Finally, we summed up both contributions in order to fit the line 
profile of Pa$\beta$ of the binary system.   
It is notable the
high velocity required to fit the primary star spectrum, viz.\, 400 km s$^{-1}$. This fact shows that the primary rotates very fast, non--synchronously and probably at critical speed.
We do not observe emission lines in the studied infrared spectral regions. 

We find transient absorption features 
around $\lambda$ 1.084, 1.093 and 1.280 $\mu$m.  
We have identified the feature around $\lambda$ 1.084 $\mu$m with He\,I 1.083 $\mu$m.
During orbital cycle this line is quite variable in strength and shape; it is stronger and wider at phases 0.33 and 0.37 and disappears in the spectrum of phase 0.21. Their RV is confusing, it follows H$\gamma$ except at phases 0.33 and 0.37 when is bluer and stronger. 
The other two absorption features are much narrower and placed at the blue side of the H\,I lines, with velocities between 550 and 850 km s$^{-1}$ from the line center. The good removal of telluric lines in the region 1.086$-$1.092 $\mu$m illustrated in  Fig.\,10 suggests that the transient features are real and not artifacts of the process of telluric correction.  While we are confident about the well defined blue--shifted narrow discrete absorption components (DACs), the quality of our data is not enough to be conclusive about the reality of some narrow
emissions features observed in some spectra.

We find DACs stronger on superphases 0.10, 0.76, 0.81 and 0.97, i.e. around long--cycle maxima, and weaker or absent on superphases 0.40$-$0.65, during long--cycle minimum (Fig.\,13). This finding could indicate that DACs are associated to the long--term periodicity. However, the correlation with their RVs is not so significant as with the orbital period (Fig.\,13).
We observe DACs with higher velocities after phases 0 and 0.5, followed by a monotonic decrease of their RVs; they trace a saw--teeth curve during every orbital cycle. 
Based on this behavior we interpret DACs as diagnostics of mass outflow from the binary. The pattern observed in Fig.\,13 can easily be interpreted as DACs arising from gas streams emanating from the external Lagrangian points $L_{2}$ and $L_{3}$. If gas accelerates through these streams following large open arcs around the binary in opposite direction to the binary motion, the observed pattern can be reproduced. Just before phase zero, we observe the weak (low density) high velocity component of the $L_{3}$ stream. Just after phase zero we observe the strong (high density) low velocity component of the $L_{2}$ stream. 

We conclude that the absorption lines observed in the infrared spectra of \var~ show a complex and variable structure, completely different to those observed in normal B--A type stellar atmospheres (Groh et al. 2007), and that our observations can be interpreted as evidence of mass loss in the binary system through the $L_{2}$ and $L_{3}$ points.

\subsection{The optical spectra, radial velocities and stellar masses}

The average SMARTS optical spectrum  shows H\,I and He\,I
absorption lines (Fig.\,14). We also detect Si\,III 4553, O\,II 4070 and O\,II 4416. 
A list of features is given in Table 6.  The
TiO bands usually found in K0$-$M2 giant/supergiant stars probably arise from the
 not resolved nearby companion included in the slit. 
The presence of O\,II 4070 lines indicates a forming region of 
high temperature and the  line strength ratio  He\,I 4387/O\,II 4416 $\approx$ 1 is indicative of high luminosity. This finding suggests the existence of a high temperature interacting region as seen in some Algol--type systems. From the inspection of the individual spectra we realized that they are highly variable: the helium and oxygen lines sometimes are visible and sometimes disappear. 
 Although helium lines and O\,II 4070  appear at all superphases, they are almost always detected during superphases 0.3$-$0.5
whereas at other superphases they may or may not appear. This 
suggests a dependence of line strength with supercycle for these lines. 
The equivalent widths of the Balmer lines change notoriously; we find that spectra taken around phase 0.5 show weaker Balmer lines that those observed around phase 0.0, consistent with a secondary star responsible for most of the H\,I absorption.  
No emission lines were detected in the optical region in these medium resolution spectra.

The high resolution MIKE spectrum shows asymmetrical H\,I absorption lines and H$\alpha$ appears weakened and with a narrow emission component
of $FWHM$ = 1.03 \AA~ at $\lambda$ = 6568.24 \AA~ (Fig.\,15). It is not clear if the H\,I lines asymmetry is due to blending of H\,I lines of the 
primary and secondary star or it is due to the presence of a central P--Cygni type absorption--emission minicomponent as suggested by the H$\alpha$ line. 
We failed to separate components with multi--gaussian deblending. 
No helium line (neither other type apart hydrogen) was observed in the MIKE spectrum.

Radial velocities of the hydrogen lines based on SMARTS spectra are displayed in Fig.\,16 along with the OGLE photometry.
To our surprise, we observe that the velocities trace the motion of the secondary; the
cooler component. 
This fact is consistent with our finding in Section 3.2 of a more luminous secondary star in the optical.
The parameters of the best fitting sinus functions are given in Table 4. 
The $\gamma$ velocity for the Balmer lines is $\approx$ 275 km s$^{-1}$ consistent with objects in the $LMC$. The radial velocities of the infrared lines, measured at minimum  of absorption profile follow the same behavior that the optical lines, but with larger amplitude 
and also  with surprisingly large $\gamma$. 
The primary contributes to the HI line motion in opposite direction to the secondary star, then $K_{2}$ should decrease, and this effect would be larger in low resolution spectra, as observed. The fact that the optical and infrared lines differs in $\gamma$ is puzzling, and cannot be due to instrumental effects, as demonstrated by the unshifted position of telluric lines in the infrared raw spectra and the consistent
optical RV with two different instrumental setups. We note that the $\gamma$ shift is quite similar for both Paschen lines, cannot be explained by line blending and is stable through several months for every line set. We cannot explain this $\gamma$ shift at present.

The mass function for a binary in a circular orbit can be expressed as:\\

$f  = \frac{M_{2}sin^{3}i}{q(1+q)^{2}} = 1.0361\times10^{-7} (\frac{K_{2}}{km s^{-1}})^{3} \frac{P_{o}}{day}$ M$_{\sun}$, \hfill(2) \\

\noindent where $K_{2}$ is the half--amplitude of the RV of the secondary star
and $P_{o}$ the orbital period.  The $f$ values derived from Eq.\,(2) and from our radial velocity study are listed in Table 4. Considering the range of $f$ and $q$, the highest possible value for $M_{2}$ is 2.80 $\pm$ 0.5 $M_{\sun}$, whereas the minimum is 0.7 $\pm$ 0.3 $M_{\sun}$.

\begin{figure}
\scalebox{1}[1]{\includegraphics[angle=0,width=8.7cm]{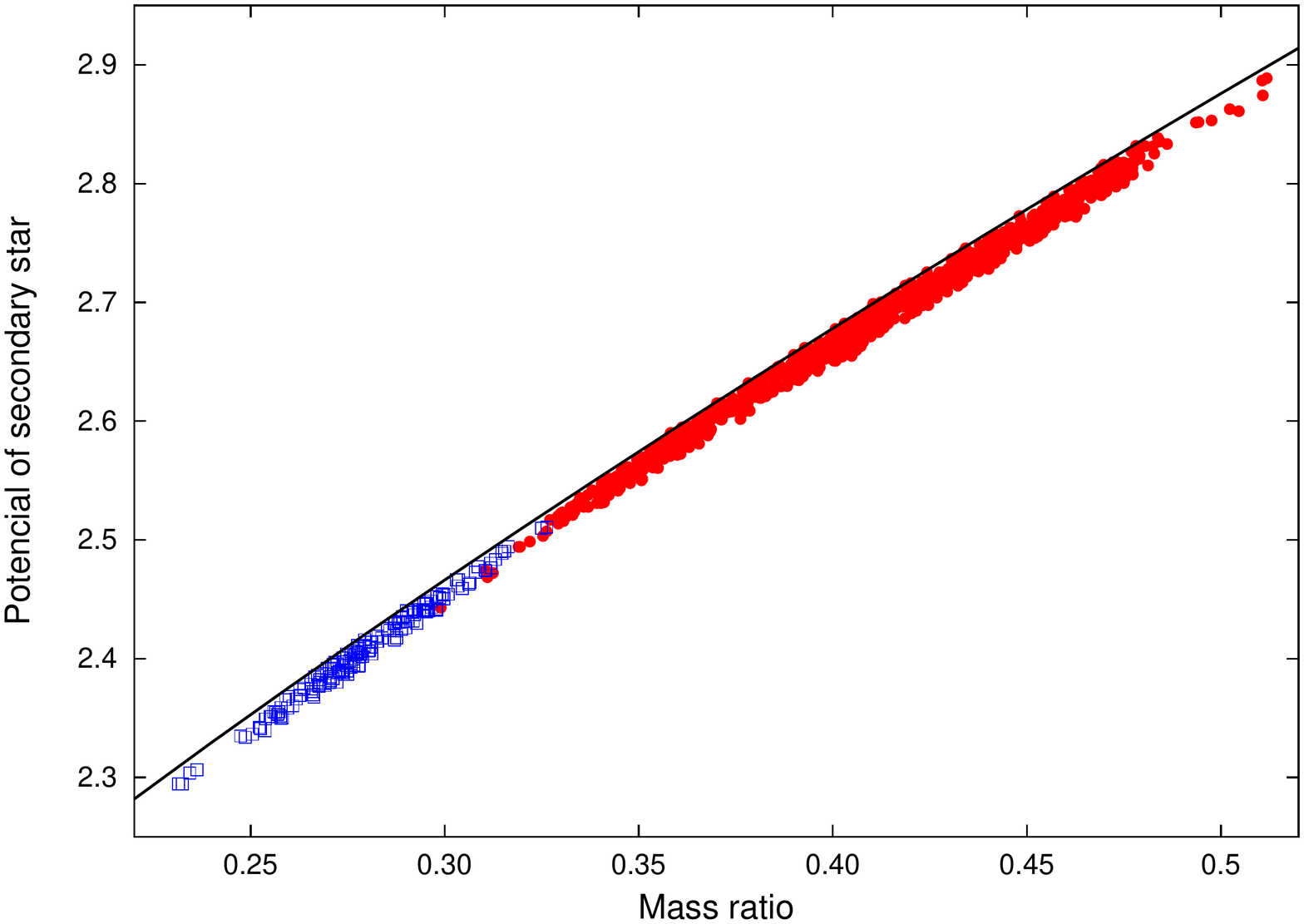}}
 \caption{Mass ratio vs. potential of the secondary component.
Only the best models are shown (the black points of previous figures). 
Parameters for OGLE fits are shown as solid circles, squares indicates MACHO blue models. 
The black line is a critical potential for object filling Roche lobe.  Our results of Monte--Carlo
simulations indicate a semi--detached configuration for \var.}
 \label{9}
\end{figure}

\subsection{Differential--slit and wide--band photometry}

We performed differential slit photometry aimed to detect the amplitude of the long--term variability
in the infrared. The observing conditions were optimal to compare the spectral flux of \var~ with their nearby companion placed simultaneously in the slit. Our observations were conducted under conditions of photometric sky, we used a wide slit and the seeing was always below 0.8\arcsec. We obtained  instrumental magnitudes in the different spectral regions defined as $-$2.5 $\log I$ where $I$ is the sum of spectral counts (ADUs) in the regions j, jplus and sz. The
central wavelengths were chosen at 1.085, 1.295 and 1.184 $\mu$m and the band widths 
0.020, 0.025 and 0.025 $\mu$m. After checking for possible 
slit misalignments and detector noise characteristics we realized that the error of 
these magnitudes is mainly due to photon noise, and neglected in all cases. When analyzing the magnitudes,
we assumed that the short--term variability in the infrared had the
same amplitude and behavior that in the $I$--band and removed this component obtaining
only the long--term infrared magnitudes. Our magnitudes are plotted along with the $I$ magnitudes in Fig.\,17.
A vertical shift was applied to minimize the scatter with the $I$ magnitudes.  Our data
suggest that there is not significant amplitude difference  between the long variability
in the $J$--band and the $I$--band.

It is difficult to make an analysis of the broad--band photometry due to the composite nature of the
spectral energy distribution (SED) and the intrinsic variability of \var. 
To have a rough idea about the SED properties we used the $UBV$ photometry by Zaritsky et al. (2004) to calculate an effective  reddening free parameter $Q$ = $-$0.526 which gives $(B-V)_{0}$ = $-$0.186 (Johnson \& Morgan 1953) and therefore $E(B-V)$ = 0.174. The unreddened effective color corresponds to
a B4\,III star (Fitzgerald 1970). The color excess is comparable with the expected in the direction of the $LMC$, namely $E(B-V) =$ 0.17 (Zaritsky 2004). 
On the other hand the infrared colors derived from Ita et al. (2004), viz.\, ($J-K$) = 0.096 and ($H-K$) = 0.076, are too red for AB--type stars 
(Koornneef 1983). 
This result is also revealed in the spectral energy distribution through multiwavelength broad--band photometry (Fig.\,18). This figure shows that the optical spectral distribution at maximum light is best represented by the 20\,000 $K$ model, roughly consistent with comparable contributions of both stellar components at this phase. The infrared data, although not necessarily obtained at maximum as 
the optical one, clearly indicates that there is a significant infrared excess. This color excess
seems to increase beyond 2.2 $\mu$m, obviously cannot be reproduced by any of the binary stellar components and probably comes from the cool circumbinary  disc responsible of the long--term variability.

Our models indicate that, at phase zero, the secondary star contributes about seven times more light that the primary in the MACHO--blue band. If we use the fact that the system has $V$ = 16.12 at minimum and assume a distance modulus for the $LMC$ of 18.39 $\pm$ 0.05 (van Leeuwen et al. 2007) and a color excess of  $E(B-V) =$ 0.17, we get $M_{V}$ = $-$2.83 for the secondary star, close to the spectroscopic estimate. Among semidetached Algols, only V\,356 Sgr (B3V+A1--2III; Budding et al. 2004, Ibano{\v g}lu et al. 2006) has a more luminous secondary. The mass--luminosity relationship for main sequence massive stars $L \sim M^{2.76}$ (Vitrichenko, Nadyozhin and Razinkova 2007) implies in our case a mass of 12.9 M$_{\sun}$. Since the dynamical mass is much lower, this result  suggests that the secondary star  could be much overluminous  and evolved. 

For a star filling their Roche lobe the mean density is constrained by the orbital period for
$q \leq$ 0.8 (e.g. Frank, King and Raine 2002):\\

$\overline{\rho} \approx 110 (\frac{P_{o}}{hr})^{-2}$ g cm$^{-3}$. \hfill(3)\\

\noindent In the case of \var~ we find $\rho$ = 3.6 $\times$ 10$^{-3}$ g cm$^{-3}$ or $\log \rho/\rho_{_{\sun}}$ = $-$2.59. 
 Approximating $R_{2}$ with  the Roche lobe radius of the secondary star (Eggleton 1983):\\

$\frac{R_{2}}{a} =\frac{0.49q^{2/3}}{0.6q^{2/3}+ln(1+q^{1/3})}$   , \hfill(4) \\

\noindent we found $R_{2}$ = 0.28$a$ or 0.30$a$, for $q$ = 0.29 and $q$ = 0.39, respectively. As the ratio between the stellar radii is $\approx$ 1.85, we get $R_{1}  \approx$ 0.16$a$.

\begin{table}
\centering
 \caption{Lines detected in the average optical spectrum of \var. The symbol $\dagger$ indicates
 probable lines of the nearby companion.}
 \begin{tabular}{@{}cccc@{}}
  \hline
$\lambda$ (\AA)& line &$\lambda$ (\AA)&line \\
\hline
3908.7&Si\,I\,3906$\dagger$&4474.4&He\,I\,4471   \\     
3935.2&Ca\,II\,3934 &4717.1&He\,I\,4713  \\          
3973.2&H$\epsilon$   &4537.4&Ti\,I+Ti\,II 4533$\dagger$  \\       
4105.4&H$\delta$     &4556.0&Si\,III\,4553   \\                 
4014.5&Ti\,I+Fe\,I 4010$\dagger$             & 4865.5&H$\beta$   \\     
4028.7&He\,I\,4026   &4924.9&He\,I\,4920  \\     
4074.6&O\,II\,4070   &  4969.1&Mn\,I\,4966$\dagger$   \\      
4146.9&He\,I\,4143 & 5003  &TiO band 5003$\dagger$ \\
4344.4&H$\gamma$  &5170& TiO band 5167$\dagger$\\  
4390.4& He\,I\,4388 & 5274.8&Ca\,I+Fe\,I 5270$\dagger$\\
4420.3&O\,II\,4416    & & \\   
\hline
\end{tabular}
\end{table}

\begin{figure}
\scalebox{1}[1]{\includegraphics[angle=0,width=9cm]{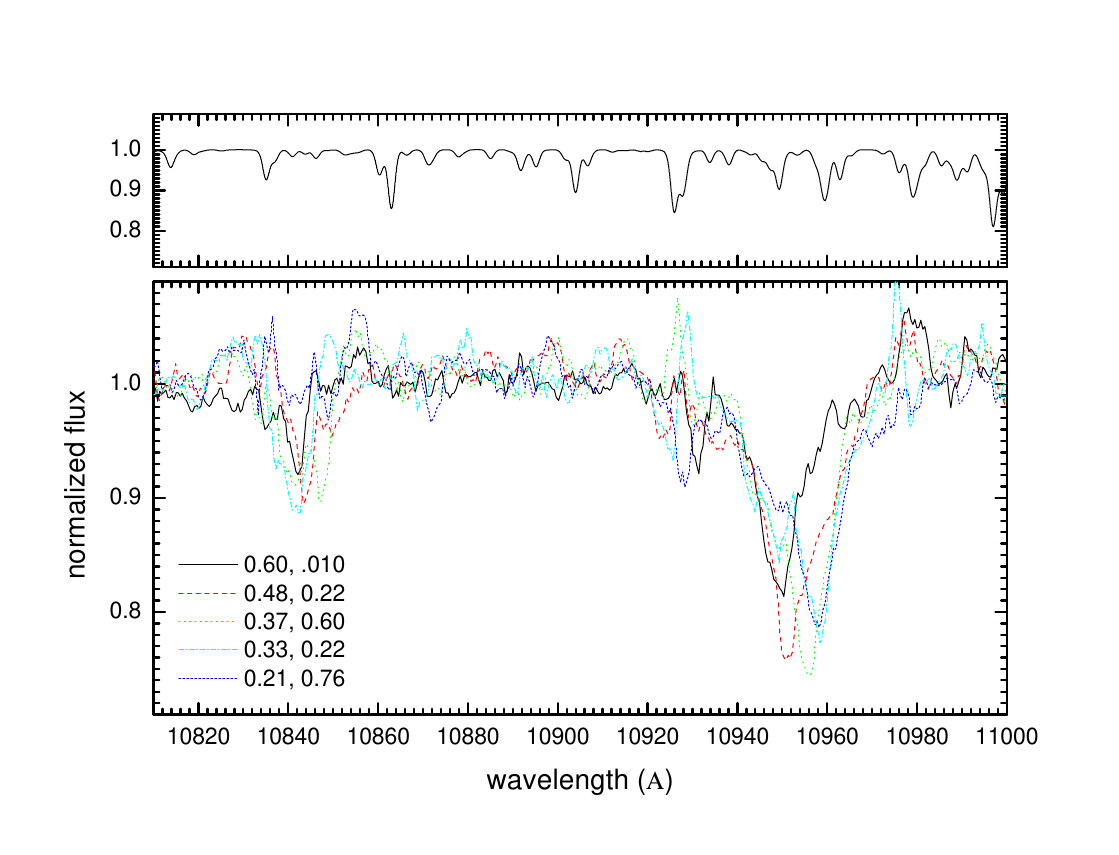}}
\caption{Pa$\gamma$ and He\,I 1.083 $\mu$m line profiles (down) and telluric template (up). 
Orbital (left) and supercycle (right) phases are indicated.
Note the discrete absorption features at the blue wing of the Pa$\gamma$ line.}
  \label{10}
\end{figure}

\begin{figure}
\scalebox{1}[1]{\includegraphics[angle=0,width=9cm]{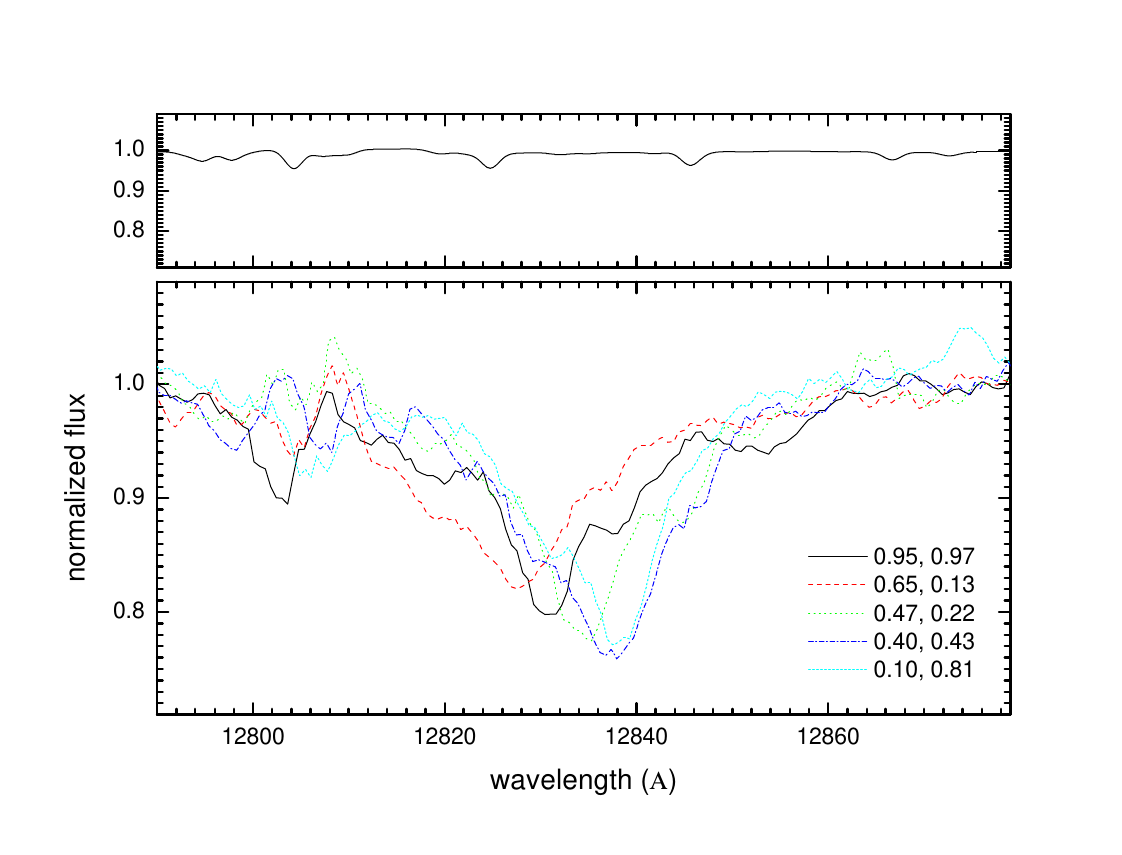}}
\caption{Pa$\beta$ line profiles (down) and telluric template (up). Orbital (left) and supercycle (right) phases are indicated.
Note the discrete absorption features at the blue wing of the Pa$\beta$ line.}
  \label{11}
\end{figure}

\begin{figure}
\scalebox{1}[1]{\includegraphics[angle=0,width=9cm]{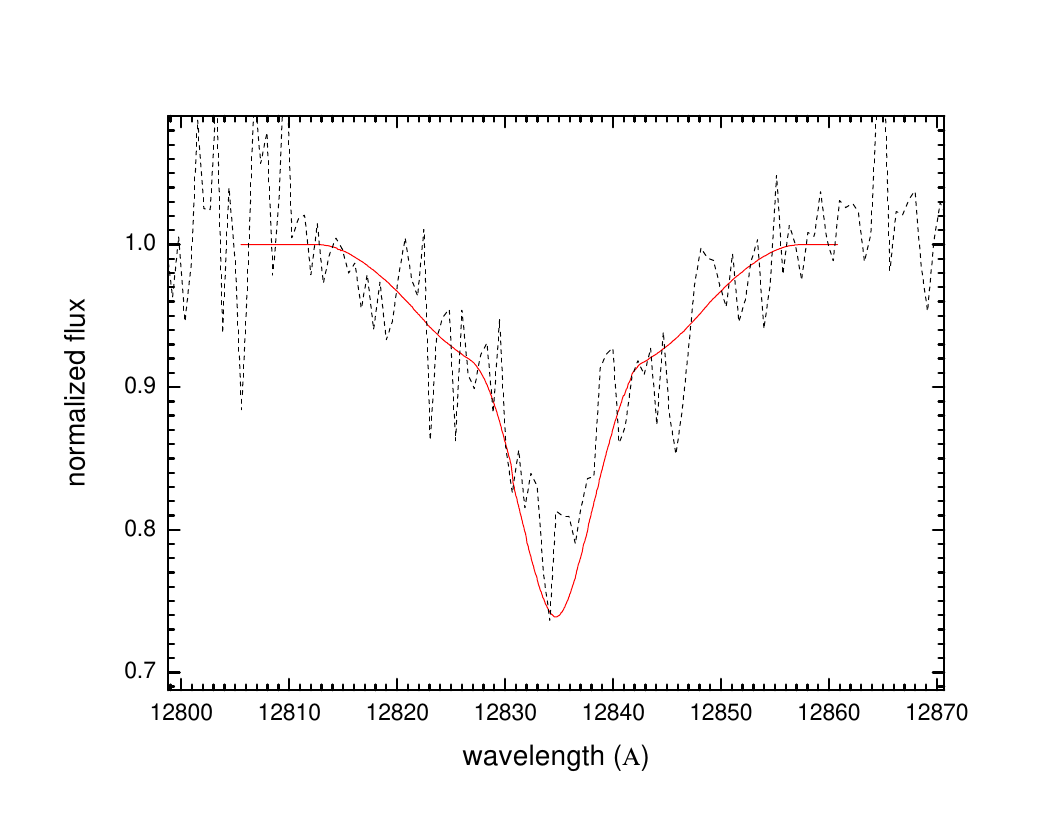}}
\caption{The Pa$\beta$ line profile at phase 0.47 (superphase 0.22) along with the synthetic binary profile built considering $vsini_{1}$ = 400 km s$^{-1}$, $vsini_{2}$ = 100 km s$^{-1}$ and the parameters given in the text.}
  \label{12}
\end{figure}

\begin{figure}
\scalebox{1}[1]{\includegraphics[angle=0,width=9cm]{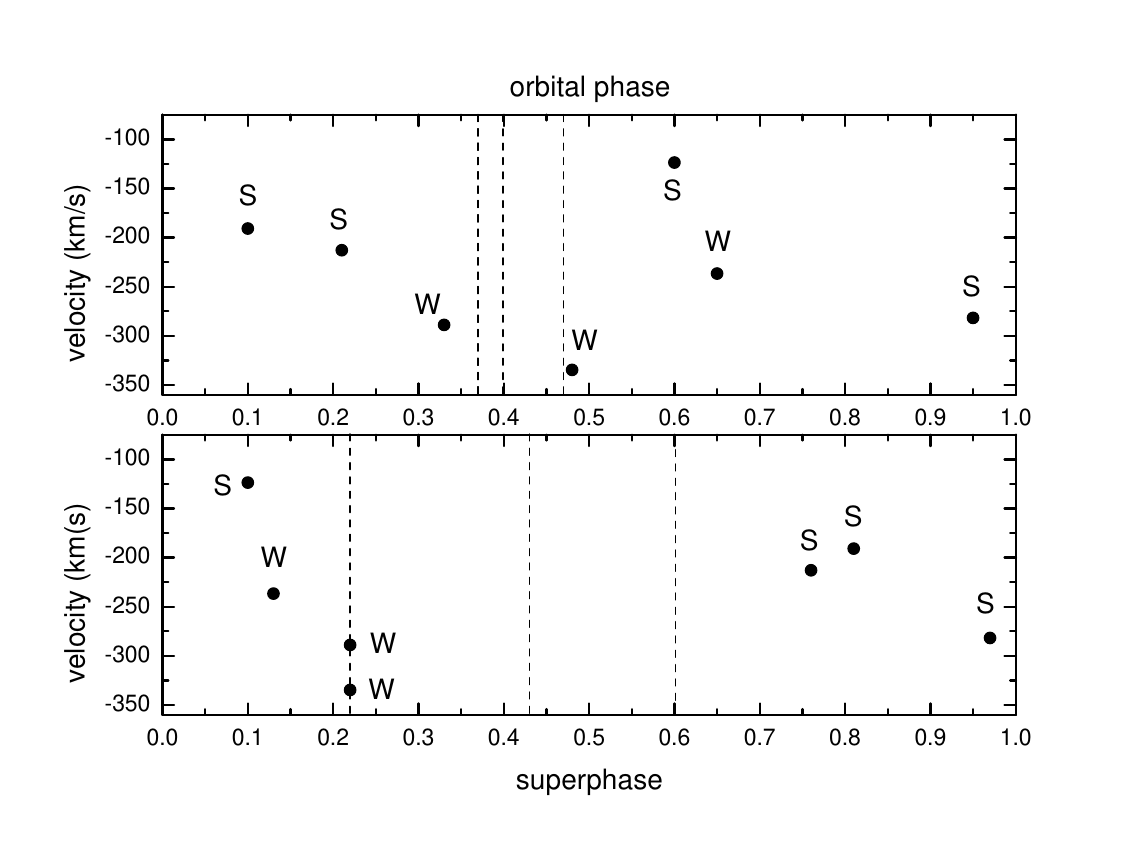}}
 \caption{Radial velocities for the discrete absorption components. We label weak (W) and strong (S) lines. Dashed lines
 indicate very weak or absent lines.  }
  \label{13}
\end{figure}

\begin{figure}
\scalebox{1}[1]{\includegraphics[angle=0,width=9cm]{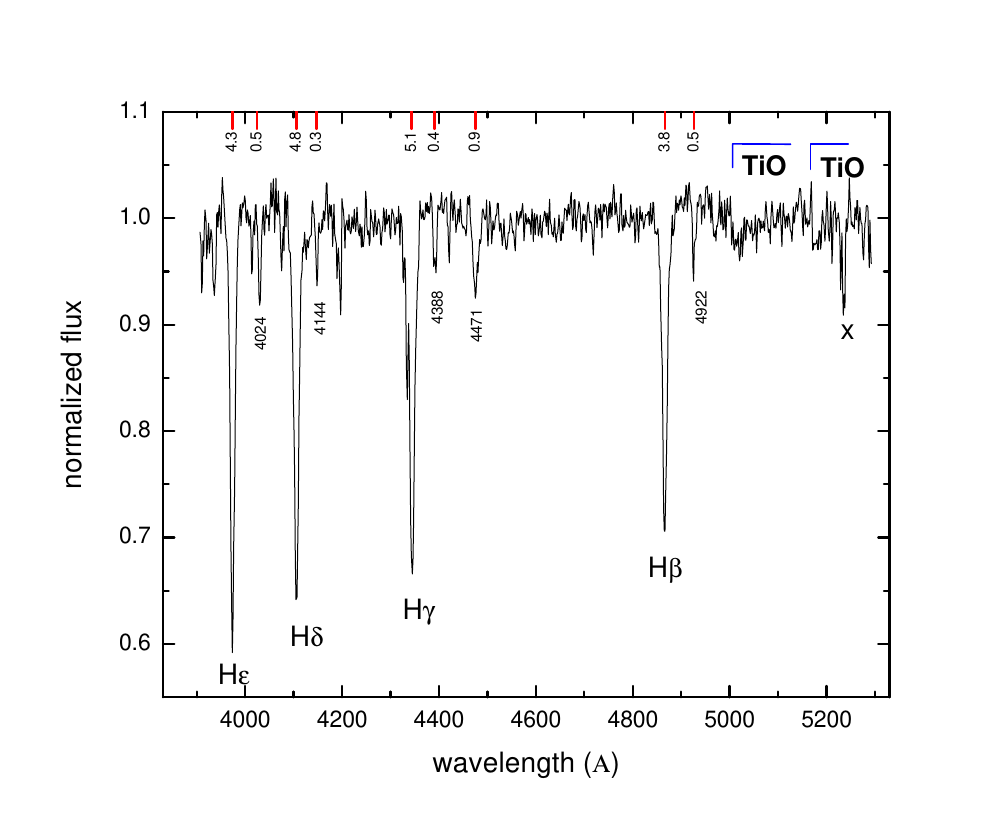}}
 \caption{Normalized average SMARTS spectrum with labels and equivalent widths (in \AA) for the main lines. The cross indicates a bad pixel region.}
  \label{14}
\end{figure}

\begin{figure}
\scalebox{.9}[.9]{\includegraphics[angle=0,width=10cm]{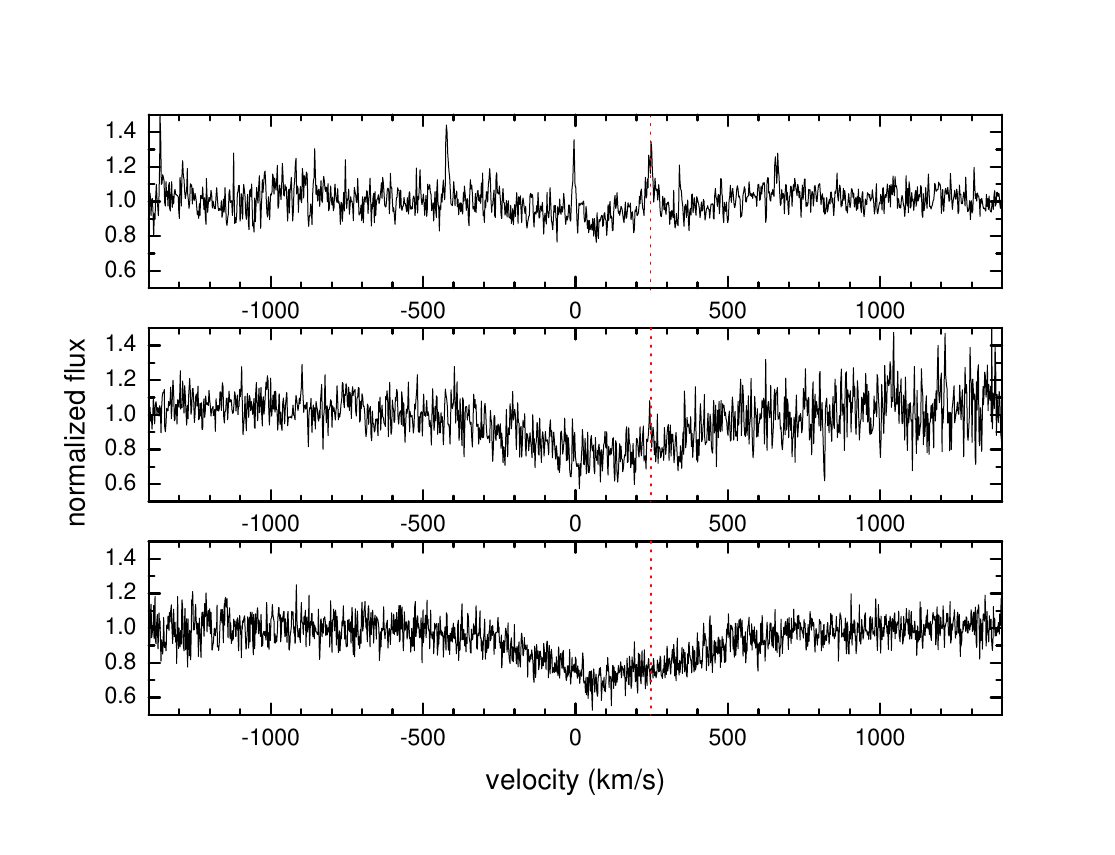}}
 \caption{Normalized high resolution MIKE spectra around H$\alpha$, H$\beta$ and H$\gamma$ (from up to down) taken at orbital phase 0.756. The velocity zeros represent the laboratory rest wavelengths.   Note the narrow emission feature inside the weakened H$\alpha$ absorption and the asymmetry of H$\beta$ and most notably H$\gamma$. The vertical dotted line indicates the velocity of the H$\alpha$ emission maximum. }
  \label{15}
\end{figure}

\begin{figure}
\scalebox{1}[1]{\includegraphics[angle=0,width=9cm]{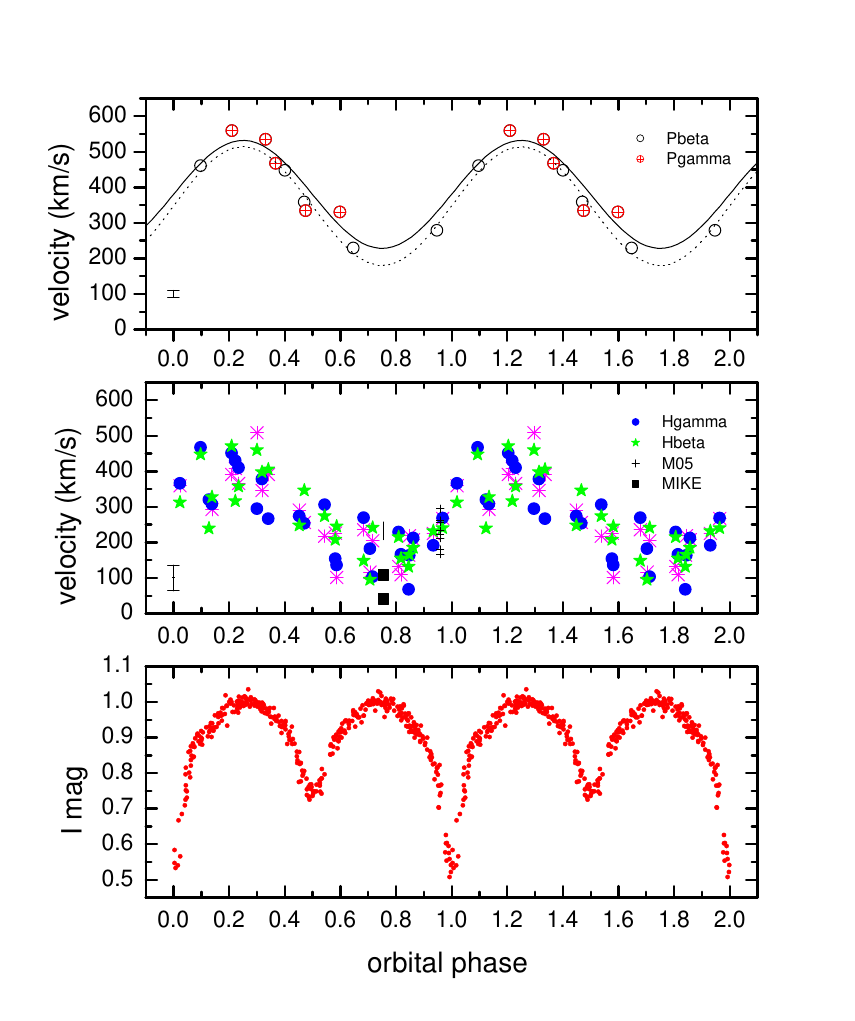}}
 \caption{Radial velocities and $I$--band magnitudes folded with the orbital period. The best sinus 
 fits are also shown. The vertical bars indicate the single measurement errors derived from the 
 $RMS$ of the standard star radial velocities. Note the $\gamma$ shift between optical and infrared lines. MIKE velocities correspond
 to the H$\gamma$ minimum and H$\gamma$ gaussian fit minimum  from the lower to the higher velocity. The velocity of the 
 peak of the H$\alpha$ emission is indicated by a vertical dash.}
  \label{17}
\end{figure}

\begin{figure}
\scalebox{1}[1]{\includegraphics[angle=0,width=9cm]{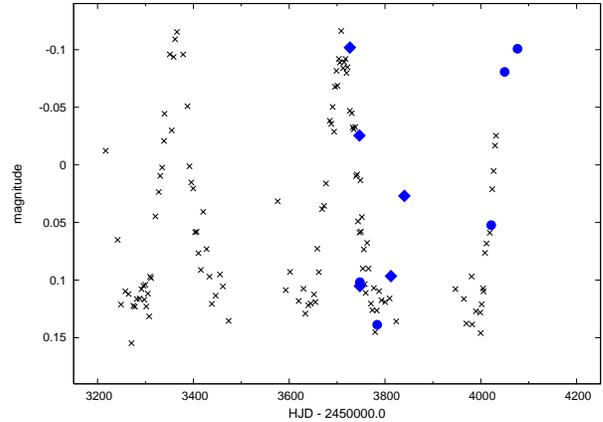}}
 \caption{Long--term OGLE\,III $I$--band light curve along with differential spectrophotometric magnitudes mj (diamonds) and mjplus (filled circles).}
  \label{18}
\end{figure}

\begin{figure}
\scalebox{1}[1]{\includegraphics[angle=0,width=9cm]{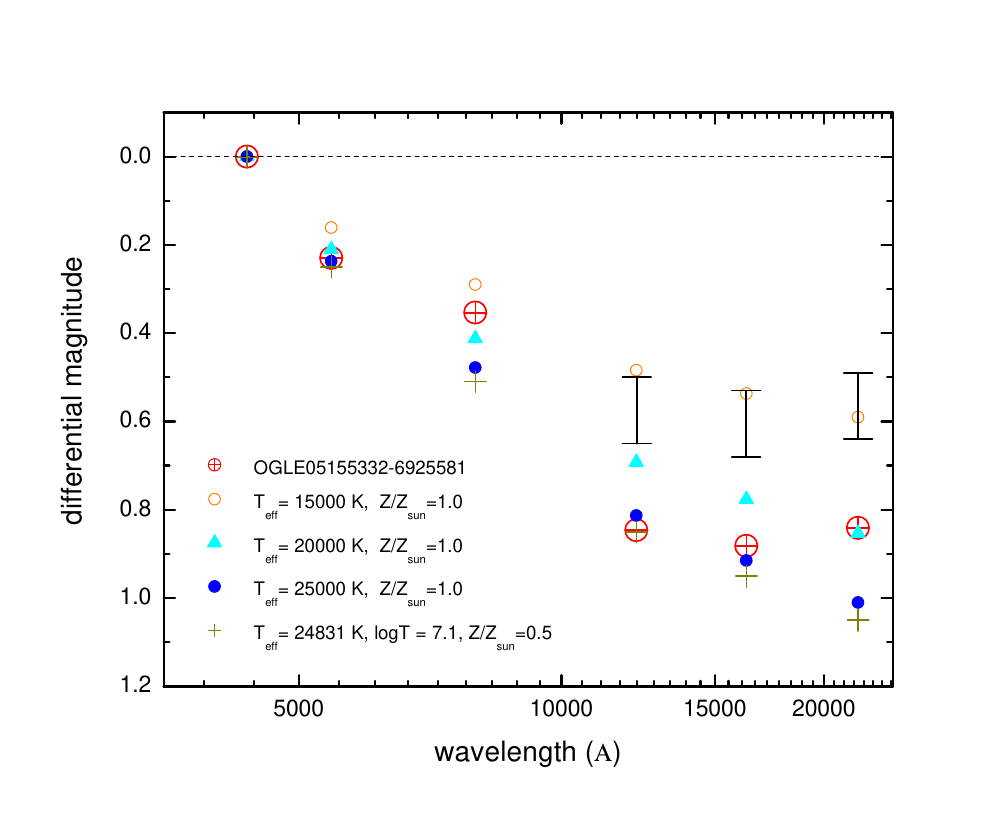}}
 \caption{Simultaneous $BVI$ magnitudes obtained at maximum and simultaneous $JHK$ magnitudes
 relatives to the $B$ magnitude. The magnitudes have been dereddened using $A_{V}$ = 0.17. We also show solar metallicity models from Bessell et al. (1998, for $\log g$ = 3.0) and  a low metallicity model from 
Bertelli et al. (1994). The bars indicate the 90\% probability loci for $JHK$ magnitudes at maximum based on 
Monte--Carlo simulations, i.e. the knowledge that they were obtained from the average of 10 randomly distributed images (Ita el at. 2004) and the assumption that the OGLE\,II light curve can be used as a seed for randomly generated infrared light curves.
The figure shows that the best model for the $BVI$ magnitudes at maximum is that with $T$ = 20\,000 $K$, and that there is a color excess already at the I--band that is prominent in redder passbands.}
  \label{19}
\end{figure}

\begin{figure}
\scalebox{1}[1]{\includegraphics[angle=0,width=9.5cm]{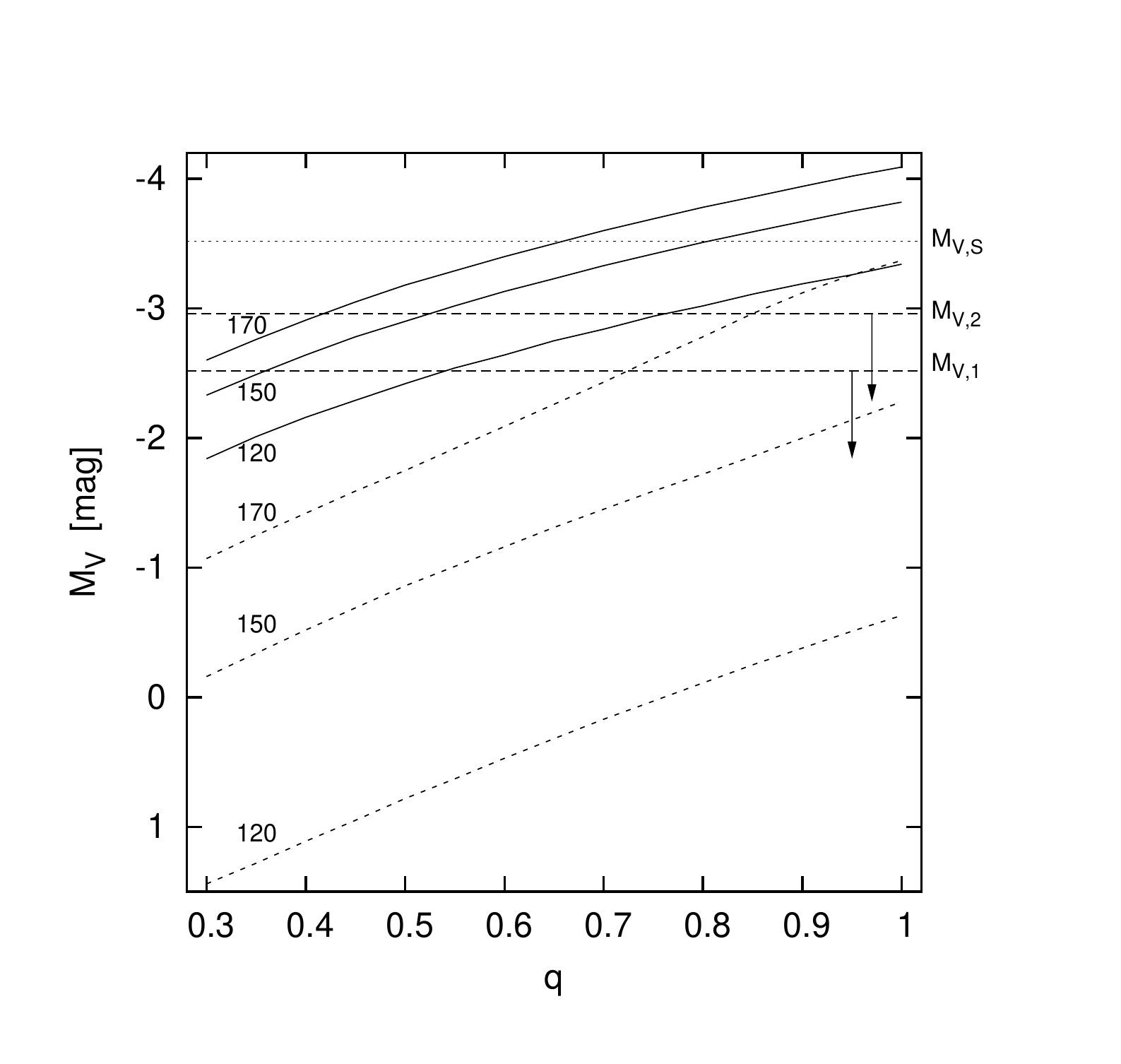}}
 \caption{ The $q$--magnitude diagram showing dynamical and model constraints for both stellar components. The horizontal lines labeled
 $M_{S},  M_{V,2}$ and $M_{V,1}$ indicate the absolute magnitude of the system, 
 secondary and primary star, respectively, for  the case of a pure stellar binary (i.e. non disc).
 In the real world these are upper limits since they include unknown contributions from the circumprimary and circumbinary discs (this uncertainty is showed  by arrows). The
 oblique dashed lines represent allowed primary star $M_{V}$ for three different  $K_{2}$ values  from Table 4. The oblique solid lines show the secondary star $M_{V}$ values for the same set of RV half amplitudes. These data comes from modeling of the secondary star as a filling Roche--lobe component with fixed $T_{2}$ =  12\,500 $K$ (lower limit). See Section 4.1 for details.}
  \label{20}
\end{figure}

\section{Discussion: proposed scenario and relevance for DPVs}

The previous analysis could indicate that \var~ is a semi--detached binary of the Algol type. 
In these binaries the less massive star appears more evolved than the more massive one, a fact that is explained if the former was initially the more massive star, evolved first and then has become the less massive component by processes of mass transfer and mass loss. The system \var~   apparently 
fulfills  this criteria, since it shows $q < 1$ and the secondary star in an advanced stage of evolution filling their Roche Lobe. This  interpretation, if confirmed, could be relevant, since
very few Algols have been found in the OGLE database of the Magellanic Clouds (Wilson 2004).

\subsection{On the circumprimary disc luminosity and mass ratio}

We have  obtained additional constraints on the mass ratio using the knowledge of the distance,  orbital period, inclination angle, half--amplitude of RV, temperature of the secondary, assumption of  spherical geometry, semidetached configuration of the system  and calculated  fractional light of the components. From the later and observed $M_{V}$ =  $-$3.52 we
estimate $M_{V,1}$ = $-$2.52 and $M_{V,2}$ = $-$2.96 as the upper limits for the stellar magnitudes. They should correspond to the stellar magnitudes in the case of pure stellar models and zero disc light contribution. Then we calculated the theoretical $M_{V,1}(M_{1}(q,K_{2},i,P),Z)$  and $M_{V,2}(\overline{\rho_{2}}(P),T_{2},M_{2}(q,K_{2},i))$ for a set of $K$ values 
assuming a main--sequence mass--luminosity relationship plus stellar evolution models (Bertelli et al. 1994)  for the first case and the mean density of a Roche--lobe filling star plus bolometric corrections given by Bessel et al. (1998) in the second case. We used $Z$ = 0.008 and $E (B-V)$ = 0.17.

Fig.\,19 shows our results in the more stringent case of the low temperature limit of $T_{2} = 12\,500 K$. To interpret the diagram we assume that the third--light contribution is neglected but not those of the circumprimary disc. In this case, it is evident that the $q >$ 1 case is rejected, as can be easily seen in the behavior of the solid  $K_{2}$ lines that cut the $M_{V,2}$ horizontal line at $q$ values lower than $\approx$ 0.75, that corresponds to the  $K_{2}$ value of 120 km s$^{-1}$.
 At this point we note that the diagram can be used  as a diagnostic for the circumprimary disc luminosity. For the case of  $K_{2}$ = 120 km s$^{-1}$ 
  and looking the corresponding dashed line we observe that the disc contribution should be about 2.5 magnitudes to the disc+primary subsystem. 
  On the other hand, if $K_{2}$ is 170 km s$^{-1}$ then $q <$ 0.4 and the disc contributions is about 1 magnitude to the disc+primary subsystem.   In all cases we observe the disc more luminous than the primary star which could explain the  general absence  of observed primary star features in the spectra. This finding cautions about the assumptions of our LC model method and its results.
  The primary eclipse could be mainly from the disc, not the primary star. In this sense, the system could be similar to $\beta\,Lyr$ (Wilson 1974). We  note that the $q < 1$ value derived here is independent  of our 
  LC modeling and robust against the presence of the circumprimary and outer discs. On the other hand, the possible non--neglectable contribution of the circumbinary disc makes the above estimated magnitudes only upper limits for the circumprimary disc contribution. In the high $K_{2}$ case, the circumprimary disc  could be comparable in luminosity with the primary star and with the circumbinary disc.

\subsection{Mass accretion and $\dot{P}$}

If the disc is the main source eclipsed during main eclipse, and its brightness is fully accretion powered and comparable in luminosity to the secondary star, then the mass accretion rate should be related to its luminosity through:\\

$L_{disc} = \frac{GM_{1}}{R_{1}}\dot{M_{1}}  \approx L_{2}$.  \hfill(5)\\

\noindent
Using an approximated mass--radius relation for main sequence stars $R_{1} = R_{\sun}(M_{1}/M_{\sun})^{0.7}$, the $M_{V}$ = $-$2.83 for the secondary star derived above and $BC$ = $-$1.2 for a 14\,000 $K$ star, we obtain $\dot{M_{1}}  \approx 1.03 \times 10^{-4}  (M_{1}/M_{\sun})^{-0.3}$ $M_{\sun}$ yr$^{-1}$, weakly dependent on the primary star mass.  For $M_{1}$ $ > $ 1.8  $M_{\sun}$ we get $\dot{M_{1}}  $ $>$  8.6 $\times 10^{-5}$ $M_{\sun}$ yr$^{-1}$. In the conservative case of mass transfer, we should observe an orbital period increase given by:\\

$\frac{\dot{P_{o}}}{P_{o}} = \frac{3\dot{M_{1}}(M_{1}-M_{2})}{M_{1}M_{2}} = \frac{3 \dot{M_{1}}(1-q)}{M_{2}}$, \hfill(6)\\

\noindent
(e.g.  Hilditch 2001).  For $q \approx$ 0.35  and $M_{2}  <$ 2.8 $M_{\sun}$ we should observe $\frac{\dot{P}}{P}$ $>$ 6  $ \times 10^{-5}$ yr$^{-1}$, being the observed upper limit 7.5 $\times 10^{-6}$ yr$^{-1}$.  We propose in Section 5 an explanation for 
the relative constancy of $P_{o}$.

\subsection{Possible evolutionary stage}

 The initially more massive star of a close binary fills their Roche lobe during one of three possible phases (Kippenhahn and Weigert 1967): during core hydrogen burning (Case A), H--shell burning
(Case B) or He--shell burning (Case C). The fact that we observe a hydrogen secondary star in \var~ would indicate that the system is the result of a Case A or Case B mass exchange.
Calculations of the evolution of a close binary component of 9 $M_{\sun}$ during Case A mass exchange show a short phase of fast mass loss (at rates $\sim$ 10$^{-3}$ $M_{\sun}$/year) from the initially more massive star lasting $\sim 6 \times 10^{4}$ years,  after which the star loses a large fraction of their initial mass, reaching 3.73 $M_{\sun}$ (Kippenhahn and Weigert 1967). During this phase the system loses matter and angular momentum through the $L_{2}$ point forming a Keplerian ring located at $\approx$ 2.25 $\times$ $a$ (Soberman et al. 1997).  As this rings carries a lot of mass and angular momentum the result is an important decrease of the orbital period. As we do not observed this rapid period change in \var,  the system is probably  after this rapid stage, in the $q < 1$ regime, consistent with our photometric mass ratio and conclusion of Section 4.1.

\subsection{On the long--term variability and circumbinary disc}

The non--eclipsing nature of the long variability places their origin outside
the binary system, probably in a circumbinary disc. It is highly likely that this disc is also responsible for the observed
infrared excess. We note that this disc is probably not the fossil of the disc formed during the phase of rapid mass loss by the binary,
since we found evidence that new material is being constantly supplied by the binary. We propose that
gas is being expelled outside the binary through the $L_{2}$ and $L_{3}$ points. The DACs pattern of Fig.\,13 strongly supports this view and indicates that the gas accelerates through the streams in open arcs around the binary in opposite directions to the orbital motion feeding the circumbinary disc.
It is possible that the remanents of earlier evolution stages form an extensive structure around the system modifying the observed properties of the object (e.g. colors). Further observations in the far infrared and radio wavelengths should be useful to check this view.

The loop encountered in the CM diagram could provide a key to understand the cause for the long term variability. 
Similar loops have been found in SMC Be stars (de Wit et al. 2006), being interpreted in terms of  
Bremsstrahlung emission arising from cycles of disc formation and dissipation around Be stars. De Wit et al. (2006)
showed that loops in the CM diagram appears when optically thick discs are formed by stellar mass ejections and then dissipate as optically thin rings into the interstellar medium. In the same way, the CM loop observed in \var~ could indicate the formation of an  outflowing optically thick circumbinary disc and their posterior dissipation in the form of an optically thin ring into the interstellar medium.  
The correlation found in DPVs between the orbital and long period (Mennickent et al.~2003) suggests that the mass loss is modulated by some kind of instability inside the binary system.

It has been shown by Packet (1981) that  during mass exchange in a 
closed binary system, enough angular momentum is transferred toward the mass accreting star to spin it up to its critical rotational velocity after gaining only a small percentage of its original mass. Hence, the primary of \var~ could be rotating at their critical velocity (see also de Mink et al. 2007). 
Our observations of a rapidly rotating primary support this view.
In this case additional gas supplied by the secondary could form a disc around the primary. The residuals observed in the LC models, the H$\alpha$ emission  and the changing He\,I lines, could be evidence of a disc plus interacting region. We propose that this disc fills the Roche lobe of the primary and escapes through the $L_{3}$ point. 
 If the system has a low mass ratio, the disc could grown beyond the 3:1 resonance radius (Lubow 1992, Murray 1996)  and precession could help to funnel material through the $L_{3}$ point (Mennickent et al.~2003). The behavior of the helium lines, which are detected primarily during long cycle minimum  shows that the interacting hot region  or its visibility is affected during episodes of strong mass loss. 

\section{Conclusions}

In this paper we have analyzed the eclipsing double periodic variable \var~ arguing that it could be a clue to understand
the phenomenon of Double Periodic Variables. We find a set of physical parameters best fitting the multiwavelength light curves and find a much improved value of the orbital period.  Our LC  and dynamical analyses indicate that \var~ is a  semi--detached binary with 
intermediate mass components where the less massive star transfers matter onto the more massive star.  We detected evidence for a luminous disc of gas around the primary as in $\beta\,Lyr$.  However,  \var~ does not show changes in the orbital period as $\beta\,Lyr$ does. In our view the system is found after a rapid mass exchange associated to Case A or Case B mass transfer, where the secondary star appears undermassive and overluminous due to the erosion of their outer layers by the mass loss process.  

We find evidence for a variable circumbinary disc, that should be the cause of the long--term periodicity and the infrared excess. Remarkably, the long--term variability follows a loop in the CM diagram (Fig.\,5) indicating quasi--cyclic episodes of disc creation/dissipation; the mass loss being modulated by some still unknown process in the binary.
The remarkable discovery of DAC RVs following a saw--teeth pattern during the orbital cycle strongly supports the view that mass is being lost from the system through accelerated streams arising from the Lagrangian $L_{2}$ and $L_{3}$ points  (Fig.\,13). This should be the source of mass for the circumbinary disc. 
  The constancy of the orbital period could indicate that the opposite effects of period increase and period decrease due to mass exchange and mass loss cancel in a long time scale. The supercycle clock could act like a safety valve, allowing the 
 mass to cumulate around the primary during a supercycle and then be  expelled away from the binary at the end of the supercycle, with almost no long--term effect on the system orbital period.
 
We conclude that \var, and probably all DPVs, are intermediate mass semidetached Algol type binaries in a evolutionary stage characterized by mass exchange, mass loss and circumbinary discs. DPVs looks more massive, bluer and more luminous than classical Algols; their study is potentially important to constraint non--conservative models of binary star evolution.

Despite of the significant advances in the knowledge of DPVs provided by the study presented in this paper,
we are certain that problems with  the complete interpretation of the phenomenon still persist. 
The complex and variable nature of \var~  requires better quality and more extensive
observational material than we have collected.
Firstly, we do not know what is the mechanism driving and regulating the long cycle variability. 
In addition, the identification 
of detected spectral features with specific places in the system and its environment
is still uncertain,  especially for the H$\alpha$ emission. We do not know how important is the role of the mass transfer and 
fast rotation of the primary star in the mechanism driving the long cycle variability. The interesting 
$\gamma$ shift observed between optical and infrared lines  of the secondary still has no explanation.  The stellar parameters for the primary obtained from our LC modeling are probably biased by the presence of a poorly known  circumprimary disc. Finally, the basic properties (mass, size, temperature) of the circumbinary disc are still unknown. The recent  discovery of bright DPVs in the Galaxy (Ko{\l}aczkowski \& Mennickent,  in prep.)
open new opportunities to our research. We are going to extend our project for intensive, multi--instrumental study
of these very promising targets. It will permit to reach the level of accuracy and time resolution essential for  constructing a comprehensive model of their activity.

On the other hand, the  Magellanic Cloud DPV sample is unique in their completeness and homogeneity. For these systems we roughly know distance, total absolute magnitude and luminosity; we will continue working on this sample to provide a sound basis of comparison for the Galactic sample with the goal of getting insights on the influence of the metallicity  on the DPV properties. We plan also make a detailed comparison with the Galactic classical Algols to enlighten the DPV--Algol connection.

\section*{Acknowledgments}
  We acknowledge the referee Dr. P.P. Eggleton for useful comments on a first version of this manuscript, especially for calling our attention on the possible 
importance of the circumprimary disc in modeling this system. We also thanks the second referee, who was on charge of this manuscript after the unexpected illness of Dr. Eggleton.
REM acknowledges financial support by Fondecyt grant 1070705. 
WG, GP and REM acknowledge financial support from the Chilean 
Center for Astrophysics FONDAP 15010003 and  from the BASAL
Centro de Astrof\'isica y Tecnologias Afines (CATA) PFB--06/2007. LC and AG acknowledges the Agencia de Promoci\'on cient\'ifica y Tecnol\'ogica for supporting this research with the grant BID 1728 OC/AR PICT 03--12720.\\
This paper utilizes public domain data obtained by the MACHO Project, jointly funded by the US Department of Energy through the University of California, Lawrence Livermore National Laboratory under contract No. W--7405--Eng--48, by the National Science Foundation through the Center for Particle Astrophysics of the University of California under cooperative agreement AST--8809616, and by the Mount Stromlo and Siding Spring Observatory, part of the Australian National University. NSO/Kitt Peak FTS data used here were produced by NSF/NOAO.\\

\bsp 
\label{lastpage}
\end{document}